\documentclass[12pt,a4paper]{article}

\usepackage[english]{babel}
\usepackage{epsfig}
\usepackage{amsmath}
\usepackage{amssymb}
\usepackage{amsbsy}
\usepackage{amsfonts}
\usepackage{array}
\usepackage{verbatim}
\usepackage{graphicx}
\usepackage{subfigure}
\usepackage{tabularx}
\usepackage{bm}
\usepackage{booktabs}
\usepackage[font=small]{caption}
\usepackage{stmaryrd}
\usepackage{mathtools}

\newcommand{\mB}{\mathcal{B}}
\newcommand{\mC}{\mathcal{C}}
\newcommand{\mG}{\mathcal{G}}
\newcommand{\mL}{\mathcal{L}}
\newcommand{\mM}{\mathcal{M}}

\newcommand{\mR}{\mathcal{R}}
\newcommand{\mS}{\mathcal{S}}
\newcommand{\mT}{\mathcal{T}}

\newcommand{\mV}{\mathcal{V}}

\newcommand{\Lij}{\mathcal{L}_{ij}}
\newcommand{\Lkk}{\mathcal{L}_{kk}}
\newcommand{\Li}{\mathcal{L}_{i}}

\newcommand{\Sij}{\mathcal{S}_{ij}}

\newcommand{\bu}{\mathbf{u}}

\newcommand{\bF}{\mathbf{F}}

\newcommand{\bmG}{\boldsymbol{\mG}}

\newcommand{\bS}{\mathbf{S}}

\newcommand{\bU}{\mathbf{U}}

\newcommand{\bq}{\mathbf{q}}

\newcommand{\bbar}{\overline}

\newcommand{\frho}{\bbar{\rho}}

\newcommand{\fp}{\bbar{p}}

\newcommand{\wt}{\widetilde}

\newcommand{\fu}{\wt{u}}
\newcommand{\fh}{\wt{h}}
\newcommand{\fe}{\wt{e}}

\newcommand{\fT}{\wt{T}}

\newcommand{\fS}{\wt{\mathcal{S}}}
\newcommand{\fSij}{\wt{\mS}_{ij}}

\newcommand{\fSkk}{\wt{\mS}_{kk}}

\newcommand{\fsigmaij}{\wt{\sigma}_{ij}}

\newcommand{\wh}{\widehat}

\newcommand{\de}{\partial}
\newcommand{\Div}{\nabla\cdot}

\newcommand{\taukk}{\tau_{kk}}
\newcommand{\tauij}{\tau_{ij}}

\newcommand{\hfu}{\breve{\fu}}
\newcommand{\hfT}{\breve{\fT}}
\newcommand{\hfrho}{\widehat{\frho}}
\newcommand{\hfS}{\breve{\fS}}

\newcommand{\bn}{\mathbf{n}}
\newcommand{\bx}{\mathbf{x}}
\newcommand{\br}{\mathbf{r}}

\newcommand{\hdelta}{\widehat{\Delta}}

\newcommand{\deltaij}{\delta_{ij}}
\newcommand{\nusgs}{\nu^{{\rm sgs}}}

\newcommand{\MA}{M\hspace{-1pt}a}
\newcommand{\PR}{Pr}

\begin{document}

\title{Direct and Large Eddy Simulation of three-dimensional non-Boussinesq gravity currents with a high order DG method}
\author{Caterina Bassi$^{(1)}$\\
 Antonella Abb\`a $^{(2)}$, Luca Bonaventura $^{(1)}$, Lorenzo Valdettaro $^{(1)}$}
\maketitle

\begin{center}
{\small
$^{(1)}$ MOX -- Modelling and Scientific Computing, \\
Dipartimento di Matematica, Politecnico di Milano \\
Via Bonardi 9, 20133 Milano, Italy\\
{\tt caterina.bassi@polimi.it, luca.bonaventura@polimi.it, lorenzo.valdettaro@polimi.it \\}
}
{\small
$^{(2)}$ Dipartimento di Scienze e Tecnologia Aerospaziali, Politecnico di Milano \\
Via La Masa 34, 20156 Milano, Italy\\
{\tt antonella.abba@polimi.it}
}
\end{center}

\date{}


\vspace*{0.5cm}


\vspace*{0.5cm}

\pagebreak


\abstract{We present results of three-dimensional Direct Numerical Simulations (DNS) and Large Eddy Simulations (LES) of turbulent gravity currents with a Discontinuous Galerkin (DG) Finite Elements method.
In particular, we consider the three-dimensional lock-exchange test case as a typical benchmark for gravity currents.
Since, to the best of our knowledge,  non-Boussinesq three-dimensional reference DNS are not available in the literature for this test case, we first perform a DNS experiment. The three-dimensional DNS allows to correctly capture the loss of coherence of the three-dimensional turbulent structures, providing an accurate description of the turbulent phenomena taking place in gravity currents. The three-dimensional DNS is then employed to assess the performance of different   LES models. In particular, we have considered the Smagorinsky model, the isotropic dynamic model and an anisotropic dynamic model. The LES results highlight the excessively dissipative nature of the Smagorinsky model with respect to the dynamic models and the fact that the anisotropic dynamic model performs slightly better with respect to its isotropic counterpart.}

\pagebreak
 
\section{Introduction}
\label{intro} \indent  

Gravity currents, arising when a heavier fluid propagates into a lighter one in a predominantly horizontal direction,
 are very common  in geophysical flows.
 In atmospheric gravity currents the density difference is typically caused by the temperature difference between the cold front and the warmer surrounding air. In oceanic flows, density differences are caused instead by salinity and temperature gradients, while in pyroclastic flows the density difference is due to the presence of suspended particles in the flow.  A comprehensive description of environmental gravity currents can be found e.g. in \cite{simpson:1997}.  
In gravity currents, the density difference between the lighter and heavier fluid can range from very small to very large. In case of small density differences, density variations in the momentum equation can be neglected in the inertia term, but retained in the buoyancy term, yielding the so called Boussinesq approximation, see e.g. \cite{gill:1982}. This approximation has been employed in most experimental and computational studies of gravity currents reported in the literature, see e.g. \cite{hartel:2000}, \cite{ozgokmen:2009}, \cite{ozgokmen:2007}. However, in several of the above listed phenomena, non-Boussinesq effects become important. 

Gravity currents appear to be a particularly interesting phenomenon from the point of view of turbulence modelling, since a wide range of interesting  phenomena arise, such as breaking internal waves and Kelvin-Helmholtz instabilities. Moreover, in gravity currents, especially in the non-Boussinesq regime, consistent density differences are present. As a consequence,  gravity currents can also be a good candidate for the validation of turbulence models for compressible flows with major density differences.
 
The main purpose of the present work is to present the first  DNS results for a gravity current benchmark in the non-Boussinesq regime. More specifically, a three-dimensional lock exchange problem
analogous to that studied in \cite{ozgokmen:2009} was simulated.
 With respect to the two-dimensional DNS already present in the literature, see e.g. \cite{birman:2005}, the three-dimensional DNS allows to effectively capture the vortex stretching and the loss of coherence of the three-dimensional coherent structures, providing more insight in the turbulent phenomena arising in gravity currents. The DNS results have then been employed for the assessment of  different subgrid LES models, including more conventional ones, such as the 
 Smagorinsky model \cite{smagorinsky:1965} and the  isotropic dynamic model \cite{germano:1991}, and less conventional ones, such as an anisotropic dynamic model \cite{abba:2001}.
 
The numerical framework chosen to implement the models is that of a Discontinuous Galerkin discretization. Such a framework allows to generalize the concept of LES filter as a projection onto the polynomial space related to the discretization, thus making it possible to apply it to arbitrary unstructured meshes. This is conceptually similar to what is done in Variational Multi-Scale (VMS) models, see e.g. \cite{hughes:2001-b}, \cite{john:2010}. 
 
The LES results confirm the excessively dissipative nature of the Smagorinsky model with respect to the dynamic models and the fact that the anisotropic dynamic model performs slightly better with respect to its isotropic counterpart. As a result, we also extend to the variable density case and to a three-dimensional configuration the findings in \cite{abba:2015}, \cite{bassi:2017} on  the importance of more complex dynamical models for subgrid modeling also in the VMS framework.

The paper is organized as follows. Section \ref{turbulence_model} provides a brief introduction of the mathematical model employed for the treatment of turbulent gravity currents, while for a more detailed description of the different subgrid turbulence models and of the Discontinuous Galerkin method we refer to the appendix \ref{section:sgsmod} and \ref{section:dgmeth}, respectively. 
The set-up of the  DNS and LES experiments is described in section \ref{setup}.
The DNS results are discussed in section \ref{num_res_dns}, while
the LES are presented in section \ref{num_res_les} and their quality assessed in terms of the corresponding DNS.
Some conclusions and perspectives for future work are presented in section \ref{conclu}.

\section{The mathematical model}
\label{turbulence_model}
We provide in this section a short overview the mathematical model we employ for the description of gravity currents. The model is based on the Navier-Stokes equations, filtered with the same procedure as in \cite{abba:2015}, \cite{bassi:2017}.  
The filtering operator, which is denoted by $\overline{\cdot},$ is in-built in the DG discretization approach
and is described in detail in appendix \ref{section:dgmeth}. Here, we only point out that the filter $\overline{\cdot}$ 
is defined as the projection onto a space of piecewise polynomial functions of degree $p,$ where 
$p$ denotes the degree of the piecewise polynomial basis functions employed by the DG method.
The choice of $p$ implicitly defines a spatial filter scale $\Delta,$  whose full definition is given in appendix \ref{section:dgmeth}.

The Favre filter operator $\wt{\cdot}$ (see e.g. \cite{garnier:2009}) is then defined implicitly by the Favre decomposition, which is given for a generic function $f$ by
\begin{equation}
\bbar{\rho f} = \frho \wt{f}.
\label{eq:favre_filter}
\end{equation}
The filtered Navier-Stokes equations  we employ can be written as:
\begin{subequations}
\label{filteq}
\begin{align}
&\de_t \frho + \de_j (\frho \fu_j) = 0 \\
&\de_t \left( \frho \fu_i \right) + \de_j \left(\frho \fu_i \fu_j\right) 
+ \de_i \fp - \de_j \fsigmaij \nonumber \\
& \qquad \qquad = - \de_j \tauij + \frho f_i \label{filteq-momentum} \\
& \de_t \left(\frho\fe\right) + \de_j \left(\frho\fh \fu_j\right) 
- \de_j \left(\fu_i \fsigmaij \right)
+ \de_j \wt{q}_j   \\
& \qquad \qquad =
- \frac{1}{(\gamma -1)\MA^2}\de_j Q_j^{{\rm sgs}}
- \frac{1}{2}\de_j \left( J_j^{{\rm sgs}} - \taukk\fu_j \right)  + \frho f_j \fu_j.  \nonumber 
 \label{filteq-energy}  
\end{align}
\end{subequations}
Here, $\fsigmaij$ and $\wt{q}_i$ are the filtered diffusive fluxes, for which the following expressions
are assumed: 
\begin{equation}
\fsigmaij = \mu \fSij^d, \qquad
\wt{q}_i = -\lambda \de_i \fT,
\label{eq:constitutive-Favre}
\end{equation} 
with $\fSij = \de_j \fu_i + \de_i \fu_j$ and $\fSij^d = \fSij -
\dfrac{1}{3}\fSkk\delta_{ij}$. 
$\tauij$, $Q_j^{{\rm sgs}}$ and $J_j^{{\rm sgs}}$ are the subgrid stress tensor, the subgrid temperature flux and the subgrid turbulent diffusion flux, respectively, whose expressions are:
\begin{subequations}
\begin{align}
\tauij & = \bbar{\rho u_i u_j} - \frho\fu_i\fu_j, \label{eqn:tauij_sgs} \\
Q_i^{{\rm sgs}} & = \bbar{\rho u_i T} - \frho\fu_i\fT =
 \frho \left( \wt{u_i T} - \fu_i\fT \right), \label{eqn:Qj_sgs} \\
J_i^{{\rm sgs}} & = \bbar{\rho u_iu_ku_k} - \frho \fu_i\fu_k\fu_k =
\frho \wt{u_i u_k u_k} - \frho\fu_i\fu_k\fu_k \nonumber \\ 
& = \tau(u_i,u_k,u_k) + 2\fu_k\tau_{ik} + \fu_i\taukk.
 \label{eqn:Jj_sgs}
\end{align}
\end{subequations}
In the last equality of equation (\ref{eqn:Jj_sgs}), the generalized central moments $\tau(u_i,u_j,u_k) = \frho\wt{u_i u_j u_k} - \fu_i\tau_{jk} 
  - \fu_j\tau_{ik} - \fu_k\tauij 
- \frho\fu_i\fu_j\fu_k$ (see \cite{germano:1992}) have been introduced. 

These subgrid terms just introduced need modeling. In this work, we have employed  the Smagorinsky model \cite{smagorinsky:1965}, the isotropic dynamic model \cite{germano:1991} and the anisotropic dynamic model 
proposed in \cite{abba:2001}. A complete description of the subgrid models is given in appendix \ref{section:sgsmod}.

\section{Set-up of the numerical  experiments and definition of diagnostic quantities}
\label{setup}

We have carried out the DNS and the LES of a three-dimensional lock exchange problem at two different Reynolds numbers ($Re=3000$ and $Re=6000$), following the experimental setting of \cite{ozgokmen:2009}  for the definition of the domain and the initial and boundary conditions. Notice that, however, we have  considered a density ratio $\gamma_r=0.7$, thus working in  a non-Boussinesq regime, contrary to what was done in \cite{ozgokmen:2009}, where simulations in the Boussinesq regime were carried out.

\begin{figure}
\centering
\includegraphics[width=0.5\textwidth]{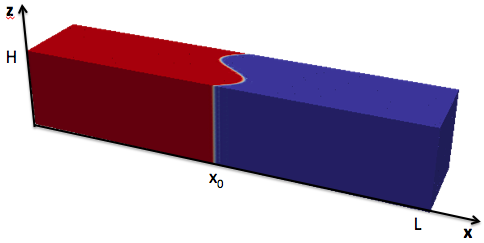}
\caption{Initial density datum for the lock-exchange test case.}
\label{fig:lock_exchange_domain_3d}
\end{figure}

The initial condition is presented in figure \ref{fig:lock_exchange_domain_3d}. The domain length is $L=5$, its height is $H=1$, its width is $W=1$, while the position of the initial discontinuity is $x_0 = 2.5$. 
The initial density datum is characterized by a sharp transition layer. Moreover, a perturbation is applied in the spanwise direction in order to ease the transition to turbulence, so that
\begin{align*}
\rho_0(x,y,z) = 
\begin{cases}
   1   & \text{if }\, 0 \leq x < a \\
   1+ 100(\gamma_r-1)(x/L + 0.495-\eta) & \text{if }\, a \leq x < b \\
   \gamma_r & \text{if }\, b \leq x \leq L, 
  \end{cases}
\end{align*}
where $a=L/2-(0.005+\eta)L$, $b=L/2+(0.005-\eta)L$, $\eta = 0.05\sin(2\pi \frac{y}{W})$ and $W$ is the width of the domain.

The initial pressure at the top of the domain is defined as in \cite{bassi:2017}:
\begin{equation}
p^{top}_{init} = \frac{1}{\gamma Ma^2},
\end{equation}
where $\gamma$ is the ratio between the specific heats and $Ma=0.1.$   An hydrostatic pressure profile is assumed in the rest of the domain. The initial temperature profile is computed from density and pressure using the equation of state for ideal gases. 
According to the non-dimensionalization employed in the present work (which is the same as in \cite{bassi:2017} and \cite{bassi:2017a}), the Froude number can be expressed as:
\begin{equation}
Fr = \sqrt{1-\gamma_r},
\end{equation}
where $\gamma_r$ is the ratio between the densities on the left and on the right with respect to the initial discontinuity. 

We employ slip boundary conditions in the streamwise and vertical directions, while periodic boundary conditions are imposed in the spanwise direction.

For the time discretization, a fourth order accurate, Strong Stability Preserving explicit Runge-Kutta method, see e.g. \cite{spiteri:2002}, has been employed for all the simulations.


In order to analyze quantitatively the DNS results and 
to assess the ability of the different LES models in reproducing them, various diagnostic quantities have been considered.
The temporal evolution of the dissipated energy was computed integrating in time the following equation:
\begin{equation}
\frac{dE_d}{dt} = \int_\Omega \left\{ \rho\nu \left[ \frac{1}{2}(\partial_j u_i + \partial_i u_j)^2-\frac{2}{3}(\nabla \cdot \textbf{u})^2\right] \right\} d\bx.
\label{eq:dissipated_energy_les}
\end{equation}
It should be noticed that the contribution of the subgrid scale viscosity $\nu_{\rm sgs}$ is added to the molecular viscosity $\nu$ in the previous equation when an LES realized with the Smagorinsky model or with the isotropic dynamic model is considered.
Since, for the anisotropic dynamic model, the isotropic and deviatoric parts of the subgrid-scale stress in the momentum equation are modeled together, the temporal evolution of the dissipated energy was computed integrating the following equation:
\begin{equation}
\frac{dE_d}{dt} = \int_\Omega \left\{ \rho\nu \left[ \frac{1}{2}(\partial_j u_i + \partial_i u_j)^2-\frac{2}{3}(\nabla \cdot \textbf{u})^2\right] - \de_i u_j \tau_{ij}^d \right\} d\bx,
\label{eq:dissipated_energy_aniso}
\end{equation}
where $\tau_{ij}^d$ is the deviatoric part of the subgrid scale stress tensor.

The second diagnostic quantity we have  considered is the time evolution of the Reference Potential Energy ($RPE$). This quantity, which has been first introduced in \cite{winters:1995}, corresponds to the minimum potential energy that can be obtained through an adiabatic redistribution of the fluid. 
Starting from the potential energy $E_p$ and the reference potential energy $RPE$, we can compute the available potential energy ($APE$) as:
\begin{equation}
APE(t) = E_p(t) - RPE(t).
\label{eq:ape}
\end{equation}
As discussed in \cite{winters:1995}, the $APE$ is the part of the potential energy that can actually be transferred to kinetic energy because of the turbulent mixing.

As pointed out in \cite{winters:1995} and recalled in \cite{ozgokmen:2009}, the reference potential energy increases with time, thanks to the stratified mixing: the $RPE$ is, as a consequence, an effective measure of how much mixing has occurred in the fluid.
We have computed the $RPE$ following the procedure outlined in \cite{tseng:2001}. We have introduced the density variable $\tilde{\rho}$ in the  sample space $[\rho_m,\rho_M]$, with $\rho_m$ and $\rho_M$ minimum and maximum densities, respectively. The probability of density $\rho$ to be in the interval $[\tilde{\rho},\tilde{\rho} + d\tilde{\rho}]$ is denoted by $P(\tilde{\rho}) d\tilde{\rho}$, where $P(\tilde{\rho})$ is a probability density function. 
This function is  estimated in practice dividing the density interval $[\rho_m,\rho_M]$ into different bins. The density field is then scanned and, for each element, we consider the density value in each Gauss integration point. If the value falls into a particular bin, a quantity equal to the Gauss weight associated to the Gauss point is accounted for that bin. The probability density function is then obtained normalizing the volume contained in each bin by the volume of the whole domain, after the completion of the whole scanning procedure.

The quantity $Z_r(\rho)$ is the height of the fluid of density $\rho$ in the minimum potential energy state, while $dZ_r$ is the thickness of the layer containing fluid of density between $\tilde{\rho}$ and $\tilde{\rho} + d\tilde{\rho}$. If the layers have the same horizontal surface $A$, the volume occupied by this layer of fluid is:
\begin{equation*}
A dZ_r|_\rho = |\Omega| P(\tilde{\rho})d\tilde{\rho} |_\rho.
\end{equation*}
This equation is then integrated over $\tilde{\rho}$ in order to obtain the profile $Z_r(\rho)$:
\begin{equation}
Z_r(\rho) = H \int_{\rho}^{\rho_M} P(\tilde{\rho}) d\tilde{\rho}.
\label{eq:vert_ref_profile}
\end{equation}
Since the $RPE$ is defined as the potential energy of the reference state, whose vertical profile is given by equation (\ref{eq:vert_ref_profile}), the following equation was employed for its computation:
\begin{equation}
RPE = \frac{1}{Fr^2} LW\int_{0}^H \rho(Z_r) Z_r dZ_r,
\end{equation} 
where $\rho(Z_r)$ is the reference density expressed as a function of the reference coordinate $Z_r$.

We have  also compared the results of the different simulations in terms of instantaneous fields at different instants of time. We have considered in particular density fields and $Q-$criterion fields. This criterion has been introduced in  \cite{hunt:1988} by identifying  a vortex as a spatial region where the Euclidean norm of the vorticity tensor dominates that of the strain rate, i.e.:
\begin{equation}
Q = \frac{1}{2}\left(\frac{1}{4} |\Omega|^2 - \frac{1}{4}|\mathcal{S}^d|^2\right) > 0,
\end{equation}  
where $\Sij^{d} = \de_{j} u_i + \de_{i} u_j - \dfrac{1}{3}(2\de_{k} u_k)\delta_{ij}$ and $\Omega_{ij} = \de_{j} u_i - \de_{i} u_j$. Notice that the $Q-$criterion allows to distinguish between pure shearing motion and the actual swirling motion of a vortex.

\section{Results of the DNS  experiments}
\label{num_res_dns}

In this section, the main results obtained from the DNS experiments are presented, to be compared in the following
section with the corresponding LES results. 

Concerning the spatial discretization, we have used as basis functions piecewise polynomials of degree  $p=4$. Notice that overintegration has been necessary in order to obtain stable simulations. Indeed, a number of integration points corresponding to exact integration of polynomials of degree $12$ has been employed. 

In table \ref{table:dofs_num_DNS} the total number of degrees of freedom associated to the DNS is presented. The total number of DOFs employed for the DNS at Reynolds $Re=3000$ is of the same order of magnitude as the one employed in \cite{ozgokmen:2009} for the $med-res1$ simulation, realized with approximately $5000000$ DOFs. In \cite{ozgokmen:2009}, two more resolved simulations, denoted by $med-res2$ and $high-res$, respectively,  were also realized, but the authors stressed that the results of the three simulations were quite similar to each other. As a consequence we have decided to perform a simulation with a spatial resolution comparable to the $med-res1$ simulation. The total number of degrees of freedom for the DNS at Reynolds $Re=6000$ is the same as the one for the DNS at $Re=3000$: this implies that for $Re=6000$ we have an underresolved DNS.  

\begin{table}
\centering
\begin{tabular}{ccc}
\toprule
 $Re=3000$ & $Re=6000$   \\
 $3800000$   &  $3800000$  \\
\bottomrule 
\end{tabular}
\caption{Number of DOFs associated to the DNS.}
\label{table:dofs_num_DNS}
\end{table}

The computational grid for our computations was built starting from a structured hexahedral mesh. Each hexahedron is divided into $N_t$ tetrahedra. The expressions for the equivalent grid spacings are:
\begin{equation}
\Delta_x = \frac{L}{N_x\sqrt[3]{N_t N_p}}, \quad \Delta_y = \frac{W}{N_y\sqrt[3]{N_t N_p}}, \quad \Delta_z = \frac{H}{N_z\sqrt[3]{N_t N_p}},
\label{eq:resolutions}
\end{equation}
where $L$, $W$ and $H$ are the length, width and height of the computational domain, respectively, $N_x$, $N_y$ and $N_z$ are the number of hexahedra in the $x$, $y$ and $z$ directions and $N_p$ is the number of degrees of freedom per element when the polynomial degree is equal to $p$.
In table \ref{table:subdivision_num_DNS} we present the values of $N_x$, $N_y$ and $N_z$ for the DNS.
Following equations (\ref{eq:resolutions}), the resolutions employed in the different directions for the DNS at $Re=3000$ and for the under-resolved DNS at Reynolds $Re=6000$ are given by:
\begin{equation}
\Delta_x^{DNS} = 0.006, \quad \Delta_y^{DNS}=0.004, \quad \Delta_z^{DNS} = 0.004.
\end{equation}
In order to verify if the DNS we are performing are well resolved, we compare these equivalent grid spacings to the non-dimensional Kolmogorov length scale. The dimensional form of the Kolmogorov length scale can be estimated as follows:
\begin{equation}
\eta^* = \left( \frac{{\nu^*}^3}{\epsilon^*}\right)^{\frac{1}{4}},
\label{eq:kolm_length}
\end{equation}
where $\nu^*$ is the dimensional kinematic viscosity and $\epsilon^*$ is the dimensional kinetic energy dissipation. Notice that the symbol $^*$ identifies dimensional quantities. Following \cite{ozgokmen:2009}, we define the kinetic energy dissipation as:
\begin{equation}
\epsilon^* = \frac{c {u_0^*}^3}{H^*},
\label{eq:kindiss}
\end{equation}
where $c$ is a constant which, in most cases, assumes values in the interval $[0.6,0.8]$, $u_0^*$ is a characteristic velocity and $H^*$ is a characteristic length, which in our case corresponds to the height of the computational domain. 
In \cite{ozgokmen:2009}, the characteristic velocity is set equal to the velocity of the gravity current head as: 
\begin{equation}
u_0^* = \frac{1}{2}\sqrt{\frac{g^* (\rho_1^*-\rho_2^*) H^*}{\rho_2^*}}, 
\end{equation}
where $\rho_1^*$ and $\rho_2^*$, with $\rho_1^* > \rho_2^*$, are the dimensional densities on the left and on the right of the initial discontinuity. 
The non-dimensionalization is carried out  in the present work using the buoyancy velocity $u_b^*$ defined as:
\begin{equation}
u_b^* = \sqrt{\frac{g^*(\rho_1^*-\rho_2^*)H^*}{\rho_1^*}}.
\label{eq:buoyancy_velocity}
\end{equation}
As a consequence, $u_0^*$ can be rewritten as:
\begin{equation}
u_0^* = \frac{1}{2}\sqrt{\frac{\rho_1^*}{\rho_2^*}} u_b^* = \frac{1}{2}\sqrt{\frac{\rho_1}{\rho_2}} u_b^*.
\label{eq:u0_velocity}
\end{equation}
We now rewrite equation (\ref{eq:kolm_length}) employing equations (\ref{eq:kindiss}) and (\ref{eq:u0_velocity}) and highlighting the non-dimensional quantities:
\begin{equation}
\eta H^* = \left( \frac{(\nu H^* u_b^*)^3}{\frac{c}{H^*}{u_0^*}^3}\right)^{\frac{1}{4}} = \left( \frac{(\nu H^* u_b^*)^3}{\frac{c}{H^*}\left(\frac{1}{2}\sqrt{\frac{\rho_1}{\rho_2}}u_b^*\right)^3}\right)^{\frac{1}{4}}.
\label{eq:kolm_length_adim}
\end{equation}
Simplifying, we get  for the non-dimensional Kolmogorov length scale:
\begin{equation}
\eta = \left( \frac{\nu^3}{\frac{c}{8(\sqrt{\gamma_r})^3}}\right)^{\frac{1}{4}},
\label{eq:kolm_length_adim_def}
\end{equation}
where, as before, $\gamma_r=\rho_2/\rho_1$.
For the kinematic viscosity,  Sutherland's law is employed, obtaining:
\begin{equation}
\nu = \frac{T^\alpha}{\rho Re}.
\end{equation} 
As a result we obtain the following values:
\begin{equation}
\eta_{3000} = 0.004, \quad \eta_{6000} = 0.002.
\end{equation} 
If we compare the employed resolutions of equation (\ref{eq:resolutions}) with the non-dimensional Kolmogorov length scales of the previous equation, we can claim that  the DNS at $Re=3000$ is quite well resolved, while the simulation at $Re=6000$ is under-resolved. While we plan repeating the DNS at $Re=6000$ at higher resolution 
as soon as computational resources are available,  in this work we will  employ it as a reference for the assessment of the different LES models, since the number of DOFs involved in these simulations is $N_{dofs} = 300000$, thus one order of magnitude smaller with respect to that of the under-resolved DNS (compare tables \ref{table:dofs_num_DNS} and  \ref{table:dofs_num_LES}).


\begin{table}
\centering
\begin{tabular}{ccc}
\toprule
$Re=3000$ & $Re=6000$   \\
$N_x=56$, $N_y=18$, $N_z=18$   &  $N_x=56$, $N_y=18$, $N_z=18$ \\
\bottomrule 
\end{tabular}
\caption{Number of hexahedra in the $x,y,z$ directions associated to the DNS.}
\label{table:subdivision_num_DNS}
\end{table}

\begin{figure}
\centering
\begin{subfigure}[]{
     \label{fig:density_iso_3000_dns_9}
     \includegraphics[trim=0cm 0cm 2cm 1cm, clip=true, totalheight=0.42\textheight]{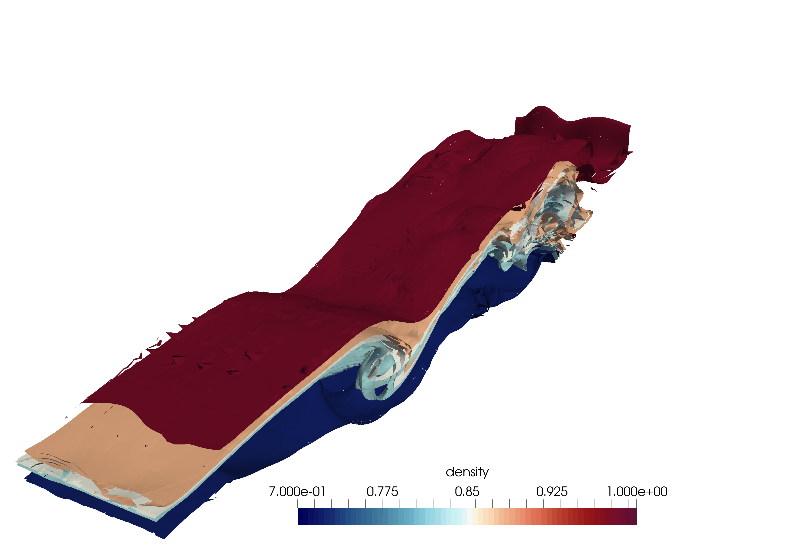}}    
\end{subfigure}

\begin{subfigure}[]{
      \label{fig:q_iso_3000_dns_9}
      \includegraphics[trim=0cm 1cm 2cm 1.4
      cm, clip=true, totalheight=0.42\textheight]{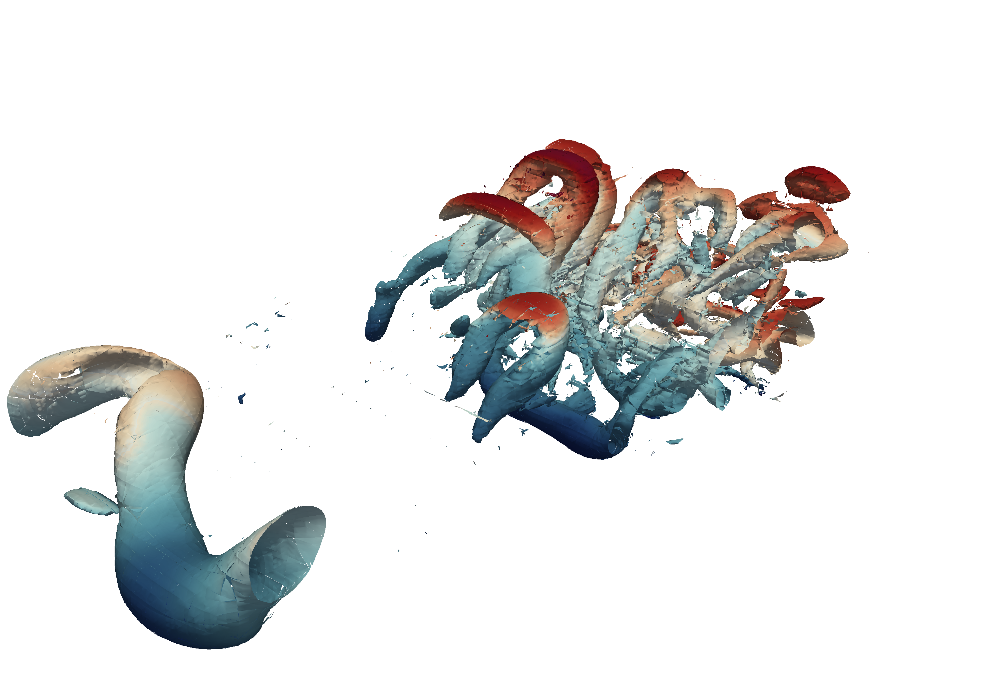}}
\end{subfigure} 
\caption{Isosurfaces at $t=9$ for $Re=3000$ (DNS) (a) density isosurfaces ($\rho = 0.72, 0.82,0.85, 0.88, 0.98$). (b) $Q=5$ isosurface (colored by density).}
\end{figure}

In figures \ref{fig:density_iso_3000_dns_9} and \ref{fig:q_iso_3000_dns_9} we show the density isosurfaces and the $Q=5$ isosurface for the $Re=3000$ DNS at $t=9$. We can see that, for this lower Reynolds number, the flow field appears quite ordered with the presence of few turbulent structures.

\begin{figure}
\centering
\begin{subfigure}[]{
     \label{fig:density_iso_6000_dns}
     \includegraphics[trim=0cm 2cm 6cm 5cm, clip=true, totalheight=0.27\textheight]{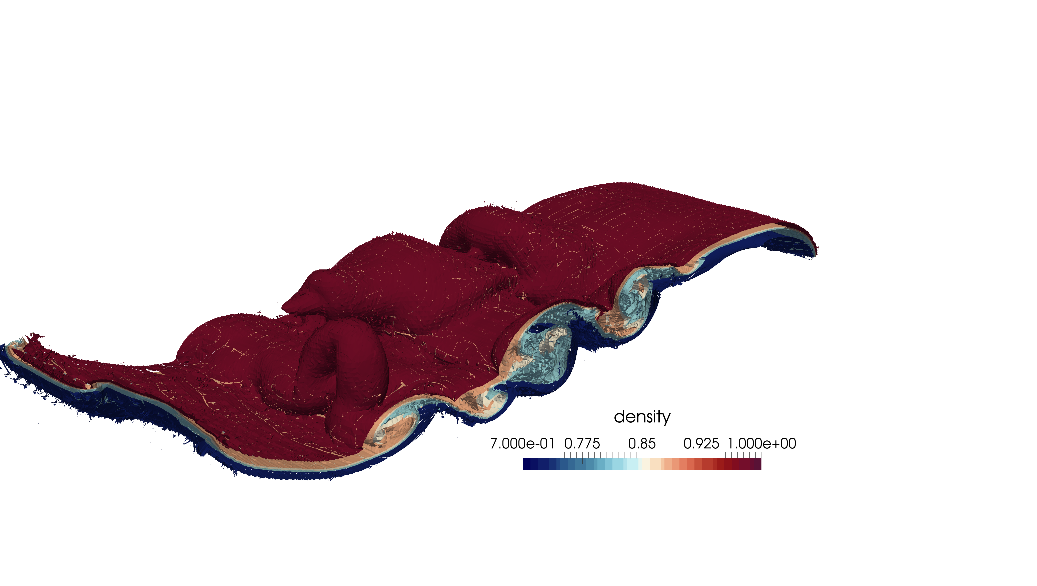}}    
\end{subfigure}

\begin{subfigure}[]{
      \label{fig:q_iso_6000_dns}
      \includegraphics[trim=0cm 0cm 4cm 2cm, clip=true, totalheight=0.35\textheight]{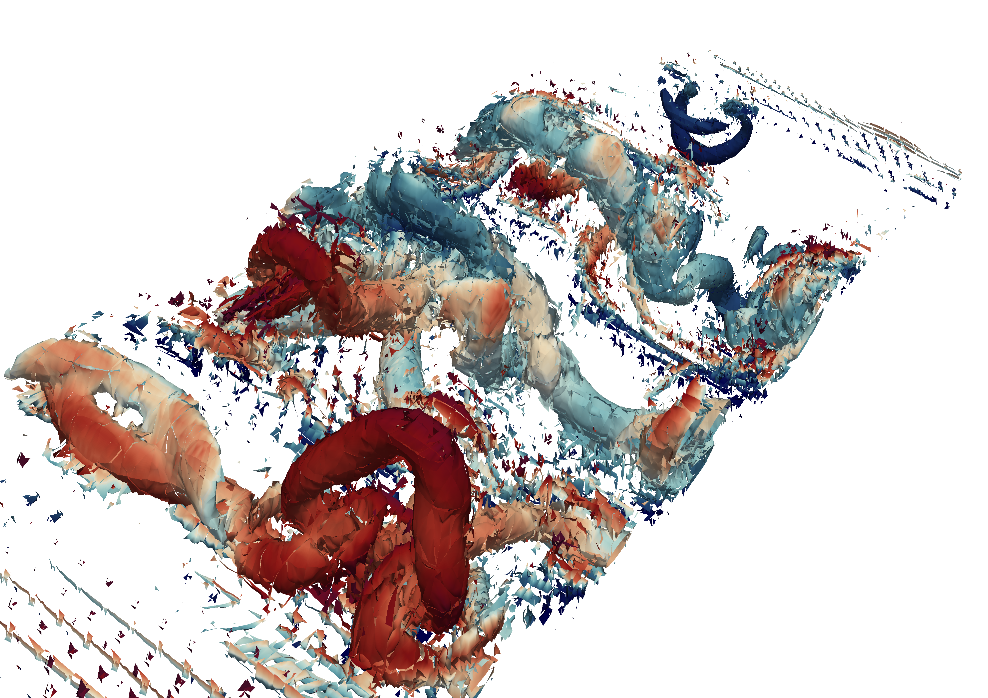}}
\end{subfigure} 
\caption{Isosurfaces at $t=4$ for $Re=6000$ (DNS) (a) density isosurfaces ($\rho = 0.72, 0.82,0.85, 0.88, 0.98$). (b) $Q=20$ isosurface (colored by density).}
\end{figure}

 In figures \ref{fig:density_iso_6000_dns} and \ref{fig:q_iso_6000_dns} we have instead 
 the density isosurfaces and the $Q=20$ isosurface 
for the $Re=6000$  DNS  at  $t=4.$  The appearance of the vortices in figure \ref{fig:q_iso_6000_dns} highlights, in this early phase of the simulation, a quite regular flow with relatively few turbulent structures even though, together with the Kelvin-Helmoltz spanwise rollers, also some longitudinal and horseshoe structures begin to appear. 

\begin{figure}
\centering
\begin{subfigure}[]{
     \label{fig:density_iso_6000_dns_9}
     \includegraphics[trim=1.8cm 1cm 2cm 1cm, clip=true, totalheight=0.42\textheight]{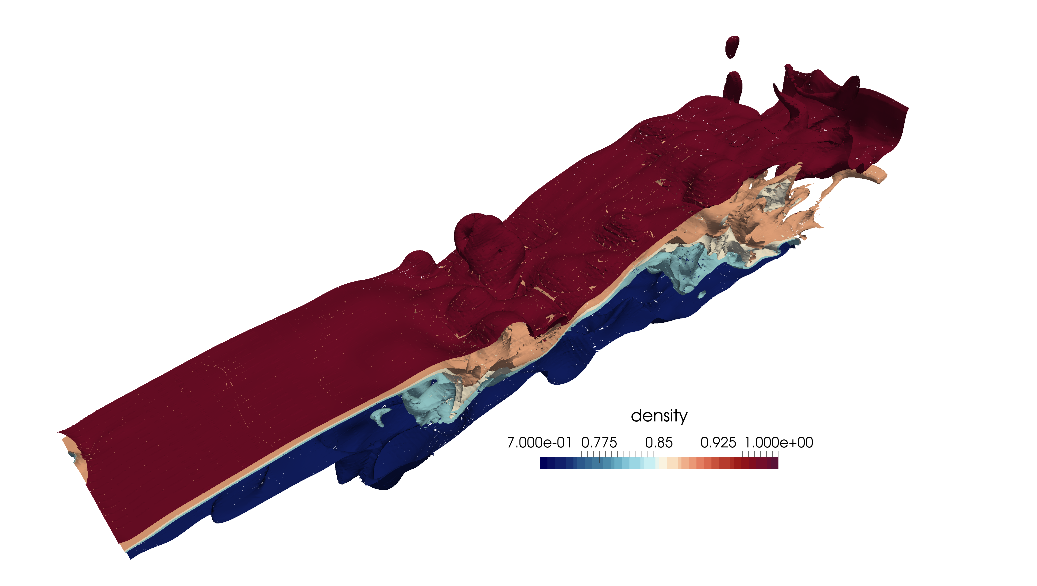}}    
\end{subfigure}

\begin{subfigure}[]{
      \label{fig:q_iso_6000_dns_9}
      \includegraphics[trim=2.6cm 1cm 2cm 1.4cm, clip=true, totalheight=0.42\textheight]{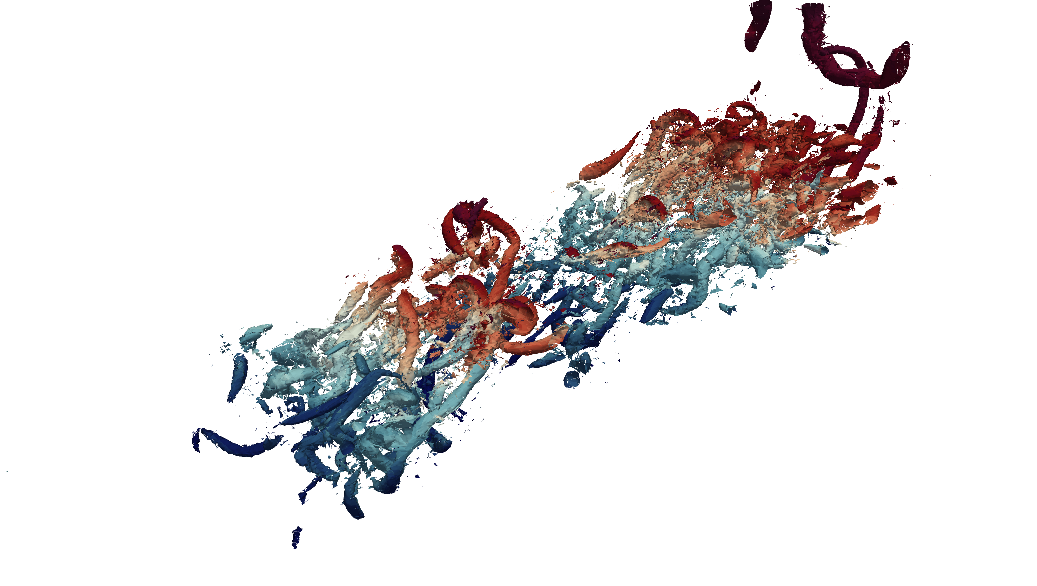}}
\end{subfigure} 
\caption{Isosurfaces at $t=9$ for $Re=6000$ (DNS) (a) density isosurfaces ($\rho = 0.72, 0.82,0.85, 0.88, 0.98$). (b) $Q=5$ isosurface (colored by density).}
\end{figure}

In figures  \ref{fig:density_iso_6000_dns_9} and \ref{fig:q_iso_6000_dns_9} we then consider the instantaneous density and $Q$ profiles for the  $Re=6000$ DNS at a more advanced instant of time, $t=9$. In particular if we look at figure \ref{fig:q_iso_6000_dns_9}, we can notice that the number of turbulent structures has considerably increased and that the flow is much more irregular with respect to the previous considered instant of time $t=4$.

\section{Results of the LES experiments}
\label{num_res_les}

In this section, the main results obtained from the LES experiments are presented and
assessed in comparison to the corresponding DNS.
 
Concerning the spatial discretization, as for the DNS, piecewise polynomials of degree $p=4$ have been employed, while the polynomial degree associated to the test filter operation, nedeed to the dynamic models, was taken to be $\wh{p}=2$. Also in the LES case overintegration has been necessary in order to obtain stable simulations.

In table \ref{table:dofs_num_LES} the number of degrees of freedom associated to the different LES is presented. The number of degrees of freedom of the LES at $Re=3000$ is of the same order of magnitude as the one employed in \cite{ozgokmen:2009} for LES at the same Reynolds number. Following  \cite{sagaut:2006}, the number of DOFs for the LES at $Re=6000$ was obtained from the number of DOFs associated to the $Re=3000$ LES as 
$$N_{dofs}^{6000} =  2^{\frac{9}{4}}N_{dofs}^{3000}.$$ 

\begin{table}
\centering
\begin{tabular}{cccc}
\toprule
$Re=3000$ & $Re=6000$   \\
$80000$   & $300000$      \\
\bottomrule 
\end{tabular}
\caption{Number of DOFs associated to the LES.}
\label{table:dofs_num_LES}
\end{table}

As for the DNS, the computational grid was built starting from a structured hexahedral mesh. 
The number of hexahedra in each spatial direction, $N_x$, $N_y$ and $N_z$, is presented in table \ref{table:subdivision_num_LES}. 
\begin{table}
\centering
\begin{tabular}{ccc}
\toprule
$Re=3000$ & $Re=6000$   \\
$N_x=15$, $N_y=5$, $N_z=5$     &  $N_x=23$, $N_y=8$, $N_z=8$     \\
\bottomrule 
\end{tabular}
\caption{Number of hexahedra in the $x,y,z$ directions associated to the LES}
\label{table:subdivision_num_LES}
\end{table}

In table \ref{table:cpu_times} we report the computational cost for the different simulations in terms of CPU hours. We can notice that the cost of the different LES is from two to three orders of magnitude smaller than the cost of the corresponding DNS. The use of a turbulence model leads to an increase of the cost of approximately one third with respect to the corresponding no-model simulation. 
\begin{table}
\centering
\begin{tabular}{cccc}
\toprule
-               & $Re=3000$ & $Re=6000$    \\
DNS             & $133900$  &  $133900$    \\
No model        & $750$     & $5400$       \\
Smagorinsky     & $875$     & $6300$       \\
Isotropic dyn.  & $1115$     & $7800$       \\
Anisotropic dyn.& $1117$     & $7800$       \\
\bottomrule 
\end{tabular}
\caption{Computational cost (CPU hours) for the different simulations.}
\label{table:cpu_times}
\end{table}
All the simulations have been performed on the Marconi cluster at CINECA. 576 cores have been employed for DNS and for the LES at $Re=6000$, while 288 cores have been employed for the LES at $Re=3000$.

\begin{figure}
\centering
\begin{subfigure}[]{
     \label{fig:density_iso_3000_smag_9}
     \includegraphics[trim=0cm 0cm 2cm 1cm, clip=true, totalheight=0.42\textheight]{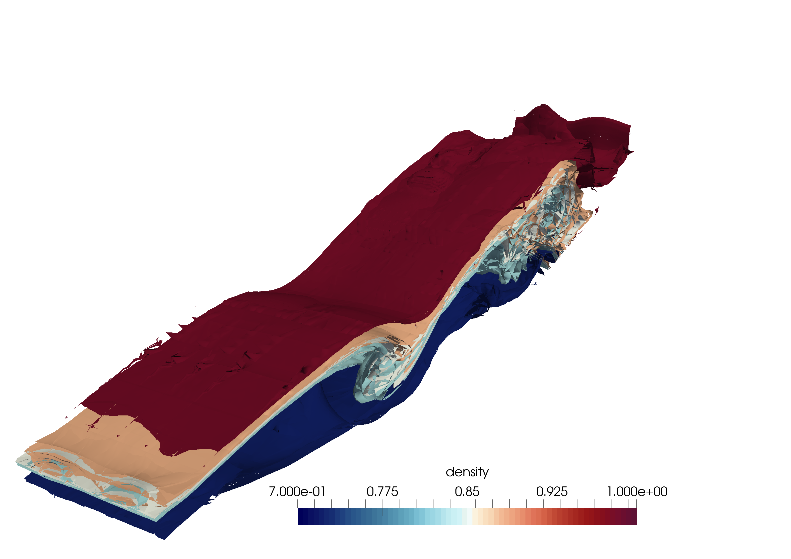}}    
\end{subfigure}

\begin{subfigure}[]{
      \label{fig:q_iso_3000_smag_9}
      \includegraphics[trim=0cm 1cm 2cm 1.4cm, clip=true, totalheight=0.42\textheight]{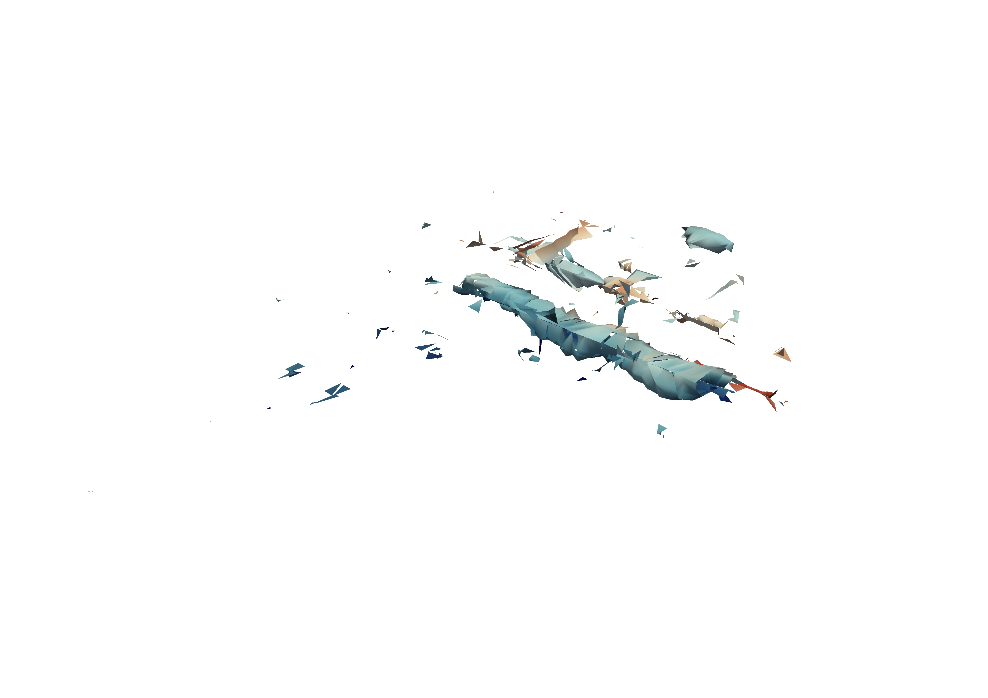}}
\end{subfigure} 
\caption{Isosurfaces at $t=9$ for $Re=3000$ (Smagorinksy model LES) (a) density isosurfaces ($\rho = 0.72, 0.82,0.85, 0.88, 0.98$). (b) $Q=5$ isosurface (colored by density).}
\end{figure}

\begin{figure}
\centering
\begin{subfigure}[]{
     \label{fig:density_iso_3000_dyn_9}
     \includegraphics[trim=0cm 0cm 2cm 1cm, clip=true, totalheight=0.42\textheight]{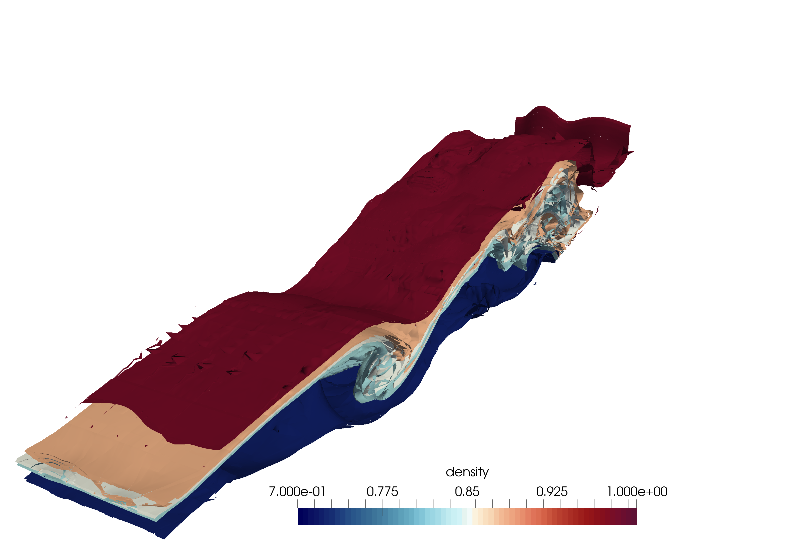}}    
\end{subfigure}

\begin{subfigure}[]{
      \label{fig:q_iso_3000_dyn_9}
      \includegraphics[trim=0cm 1cm 2cm 1.4cm, clip=true, totalheight=0.42\textheight]{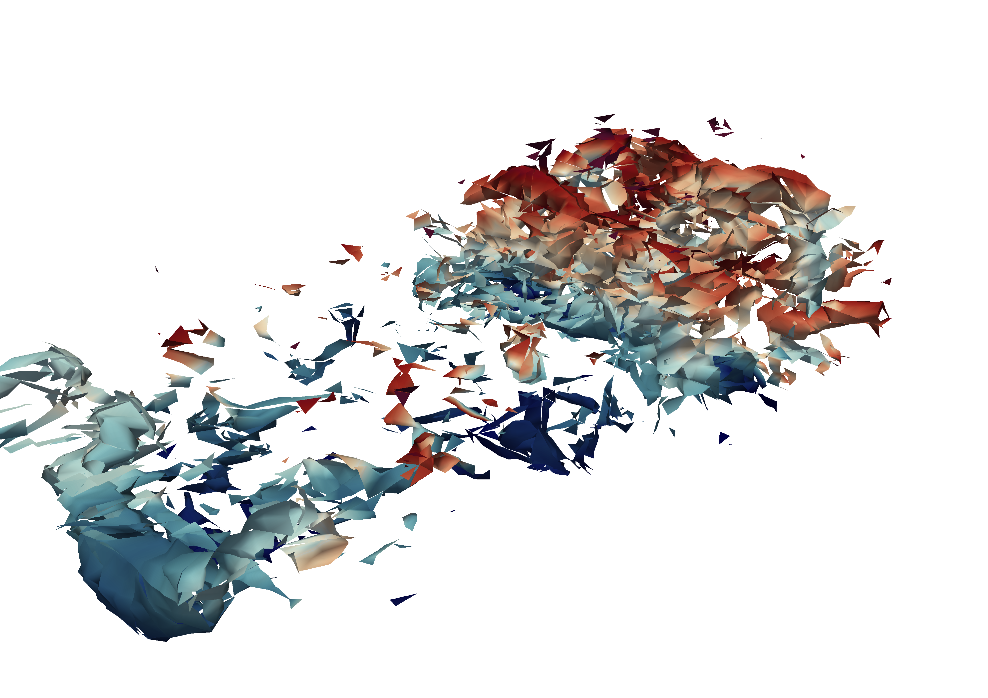}}
\end{subfigure} 
\caption{Isosurfaces at $t=9$ for $Re=3000$ (Isotropic dynamic model LES) (a) density isosurfaces ($\rho = 0.72, 0.82,0.85, 0.88, 0.98$). (b) $Q=5$ isosurface (colored by density).}
\end{figure}

\begin{figure}
\centering
\begin{subfigure}[]{
     \label{fig:density_iso_3000_aniso_9}
     \includegraphics[trim=0cm 0cm 2cm 1cm, clip=true, totalheight=0.42\textheight]{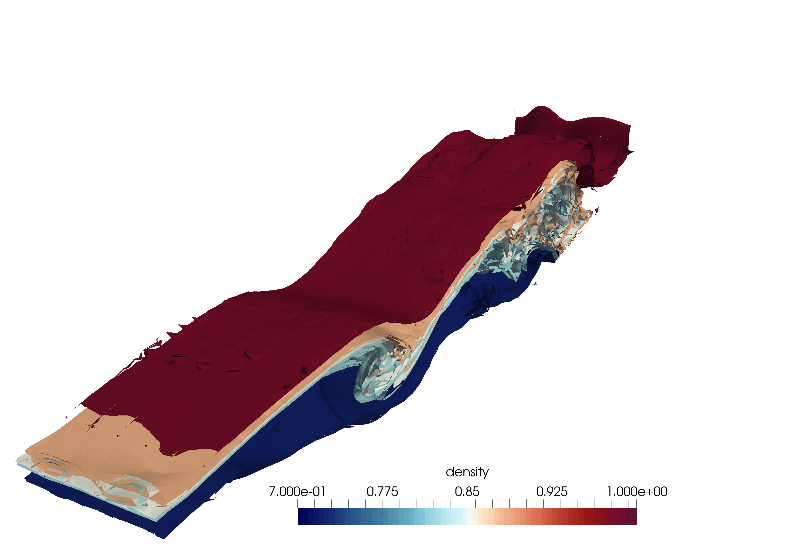}}    
\end{subfigure}

\begin{subfigure}[]{
      \label{fig:q_iso_3000_aniso_9}
      \includegraphics[trim=0cm 1cm 2cm 1.4cm, clip=true, totalheight=0.42\textheight]{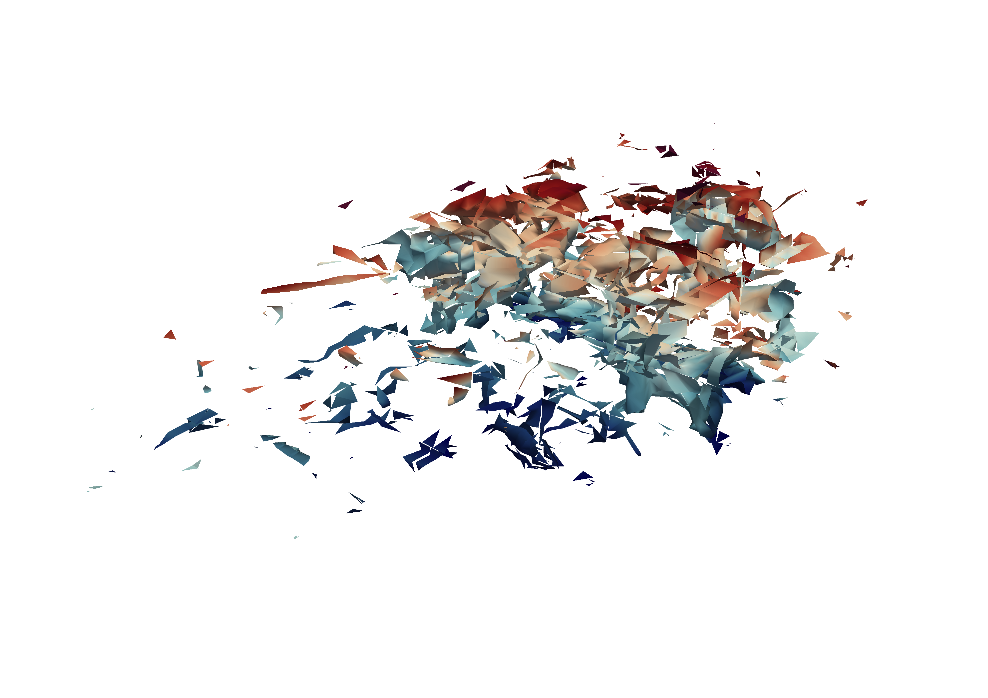}}
\end{subfigure} 
\caption{Isosurfaces at $t=9$ for $Re=3000$ (Anisotropic dynamic model LES) (a) density isosurfaces ($\rho = 0.72, 0.82,0.85, 0.88, 0.98$). (b) $Q=5$ isosurface (colored by density).}
\end{figure}

\begin{figure}
\centering
\begin{subfigure}[]{
     \label{fig:density_iso_3000_nomod_9}
     \includegraphics[trim=0cm 0cm 2cm 1cm, clip=true, totalheight=0.42\textheight]{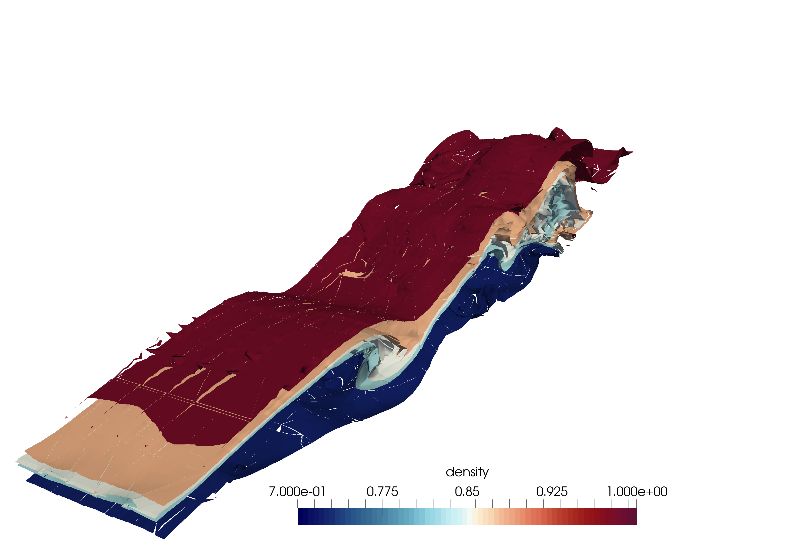}}    
\end{subfigure}

\begin{subfigure}[]{
      \label{fig:q_iso_3000_nomod_9}
      \includegraphics[trim=0cm 1cm 2cm 1.4cm, clip=true, totalheight=0.42\textheight]{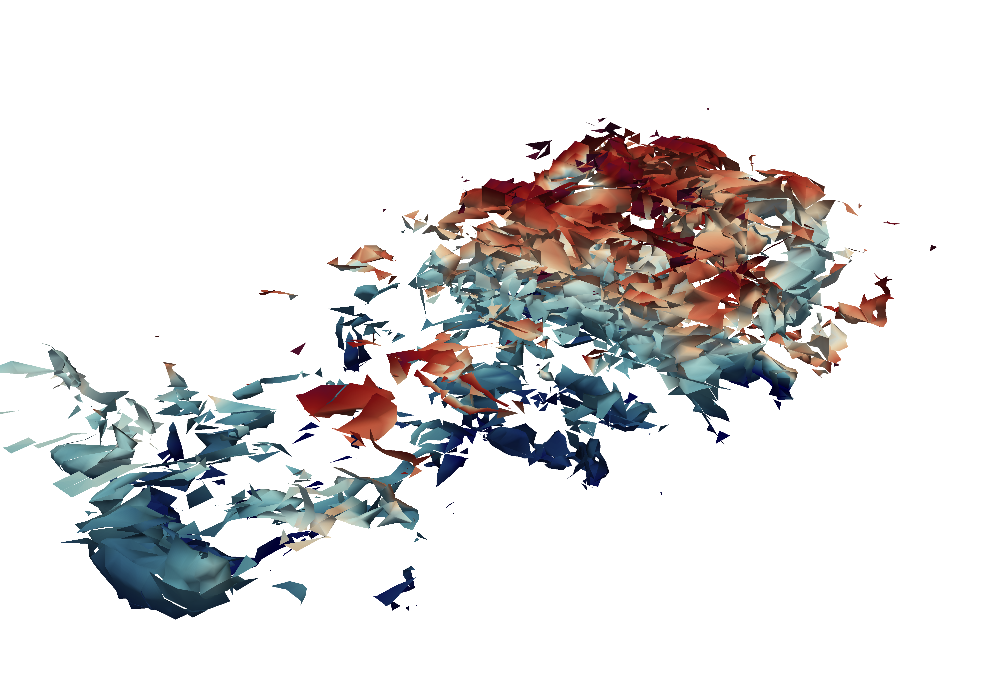}}
\end{subfigure} 
\caption{Isosurfaces at $t=9$ for $Re=3000$ (No model LES) (a) density isosurfaces ($\rho = 0.72, 0.82,0.85, 0.88, 0.98$). (b) $Q=5$ isosurface (colored by density).}
\end{figure}

Considering the $Re=3000$ case, the comparison between the different LES density fields (figures \ref{fig:density_iso_3000_smag_9}, \ref{fig:density_iso_3000_dyn_9}, \ref{fig:density_iso_3000_aniso_9}, \ref{fig:density_iso_3000_nomod_9}) and the density field of the DNS (figure \ref{fig:density_iso_3000_dns_9}),
 highlights the excessively dissipative behaviour of the Smagorinsky model. This fact is confirmed also by the comparison in terms of  the $Q=5$ isosurface. In particular we can see that the Smagorinsky model (figure \ref{fig:q_iso_3000_smag_9}) and, to a lesser extent, also the anisotropic model (figure \ref{fig:q_iso_3000_aniso_9}) provide less turbulent structures with respect to the ones provided by the corresponding DNS (figure \ref{fig:q_iso_3000_dns_9}).

\begin{figure}
\centering
\begin{subfigure}[]{
     \label{fig:density_iso_6000_smag}
     \includegraphics[trim=0cm 2cm 6cm 5cm, clip=true, totalheight=0.27\textheight]{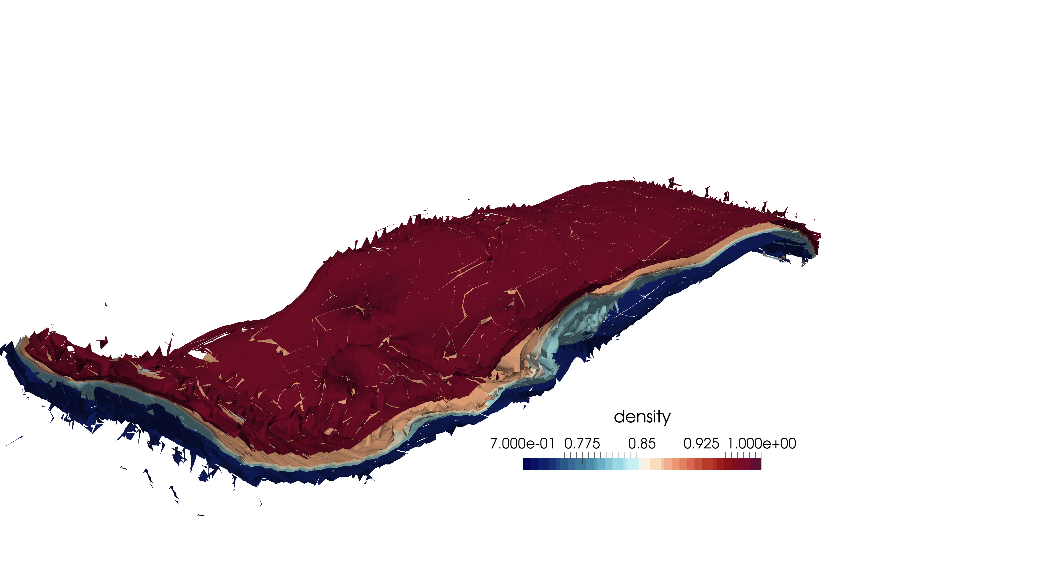}}    
\end{subfigure}

\begin{subfigure}[]{
      \label{fig:q_iso_6000_smag}
      \includegraphics[trim=0cm 0cm 4cm 2cm, clip=true, totalheight=0.35\textheight]{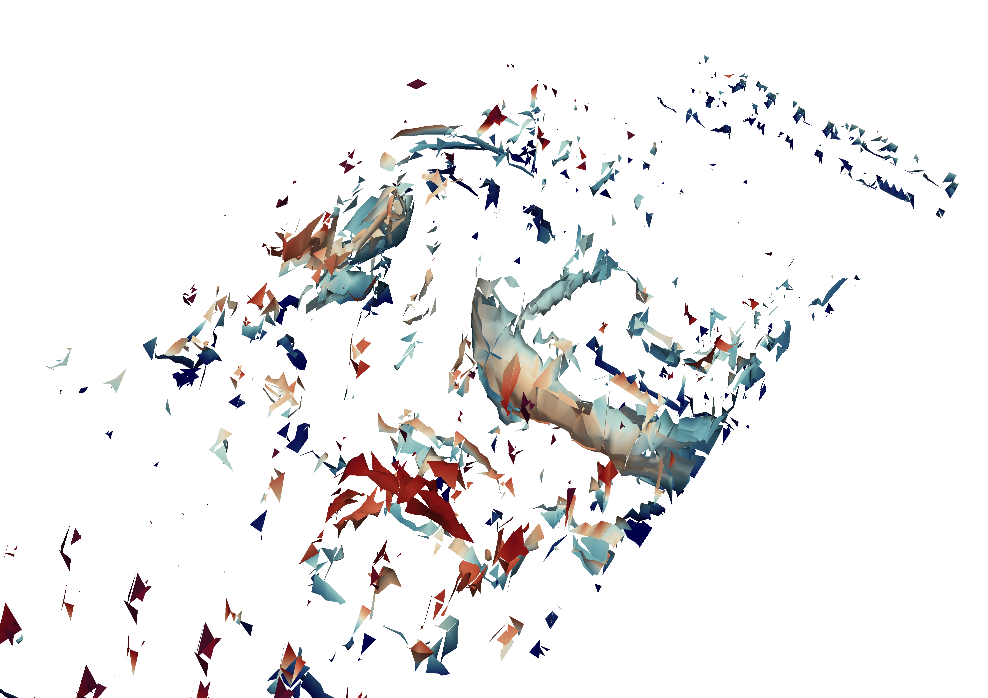}}
\end{subfigure} 
\caption{Isosurfaces at $t=4$ for $Re=6000$ (Smagorinsky model LES) (a) density isosurfaces ($\rho = 0.72, 0.82,0.85, 0.88, 0.98$). (b) $Q=20$ isosurface (colored by density).}
\end{figure}

\begin{figure}
\centering
\begin{subfigure}[]{
     \label{fig:density_iso_6000_dyn}
     \includegraphics[trim=0cm 2cm 6cm 5cm, clip=true, totalheight=0.27\textheight]{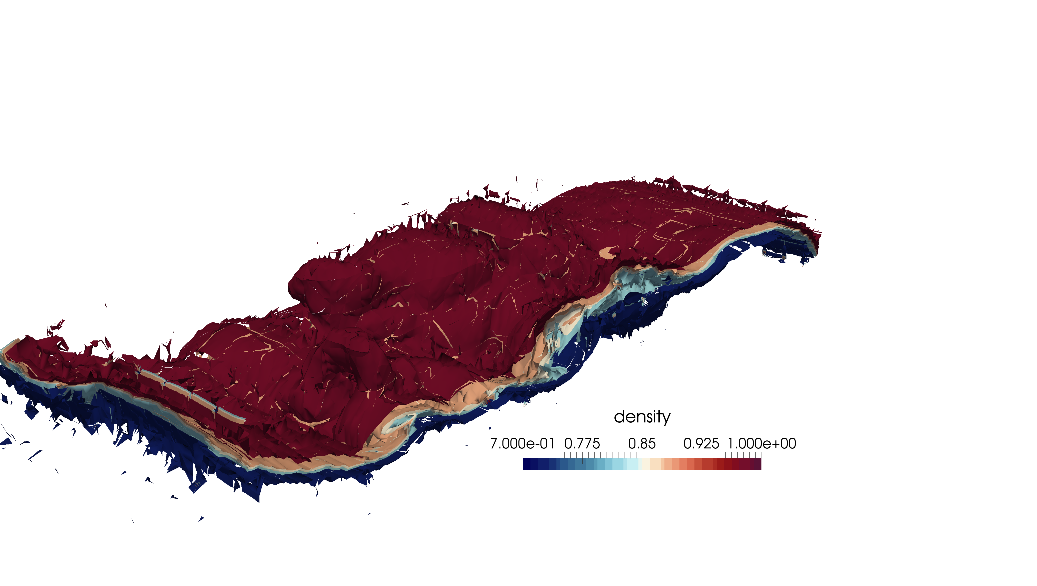}}    
\end{subfigure}

\begin{subfigure}[]{
      \label{fig:q_iso_6000_dyn}
      \includegraphics[trim=0cm 0cm 4cm 2cm, clip=true, totalheight=0.35\textheight]{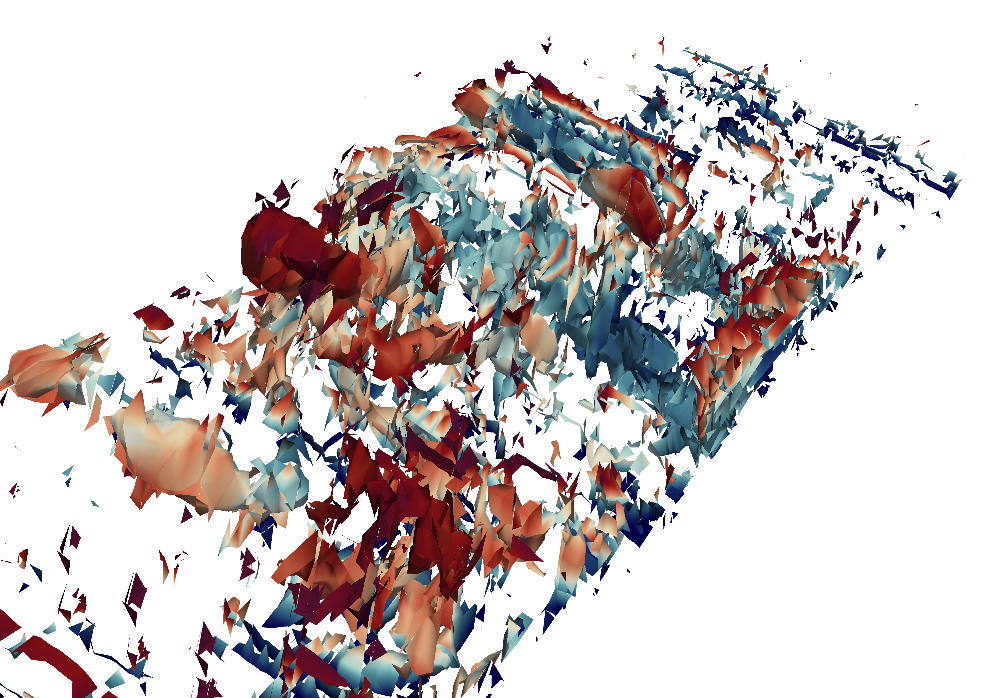}}
\end{subfigure} 
\caption{Isosurfaces at $t=4$ for $Re=6000$ (Isotropic dynamic model LES) (a) density isosurfaces ($\rho = 0.72, 0.82,0.85, 0.88, 0.98$). (b) $Q=20$ isosurface (colored by density).}
\end{figure}

\begin{figure}
\centering
\begin{subfigure}[]{
     \label{fig:density_iso_6000_aniso}
     \includegraphics[trim=0cm 2cm 6cm 5cm, clip=true, totalheight=0.27\textheight]{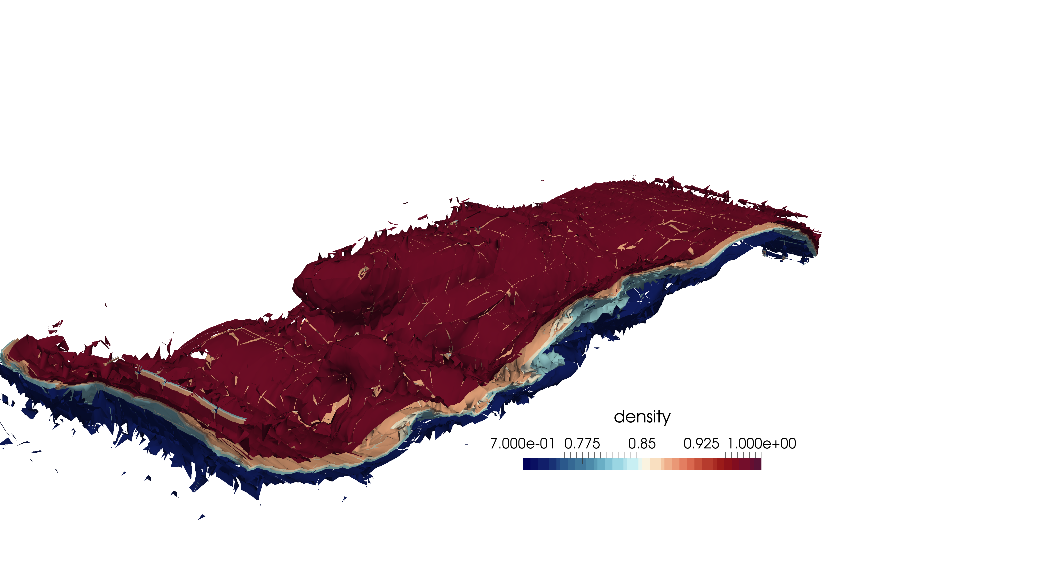}}    
\end{subfigure}

\begin{subfigure}[]{
      \label{fig:q_iso_6000_aniso}
      \includegraphics[trim=0cm 0cm 4cm 2cm, clip=true, totalheight=0.35\textheight]{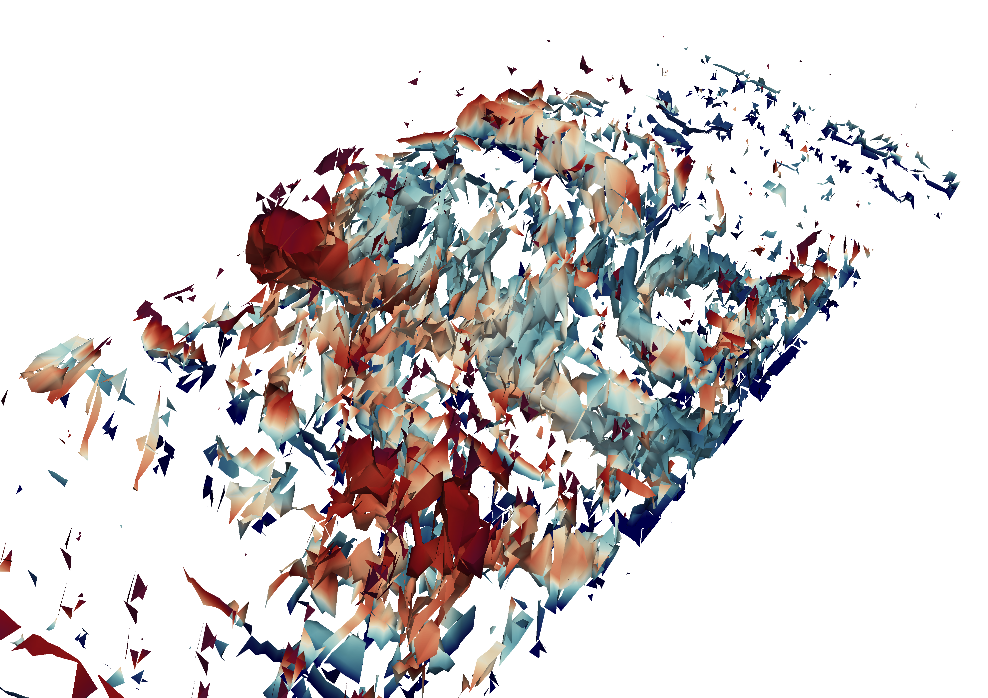}}
\end{subfigure} 
\caption{Isosurfaces at $t=4$ for $Re=6000$ (Anisotropic dynamic model LES) (a) density isosurfaces ($\rho = 0.72, 0.82,0.85, 0.88, 0.98$). (b) $Q=20$ isosurface (colored by density).}
\end{figure}

\begin{figure}
\centering
\begin{subfigure}[]{
     \label{fig:density_iso_6000_nomod}
     \includegraphics[trim=0cm 2cm 6cm 5cm, clip=true, totalheight=0.27\textheight]{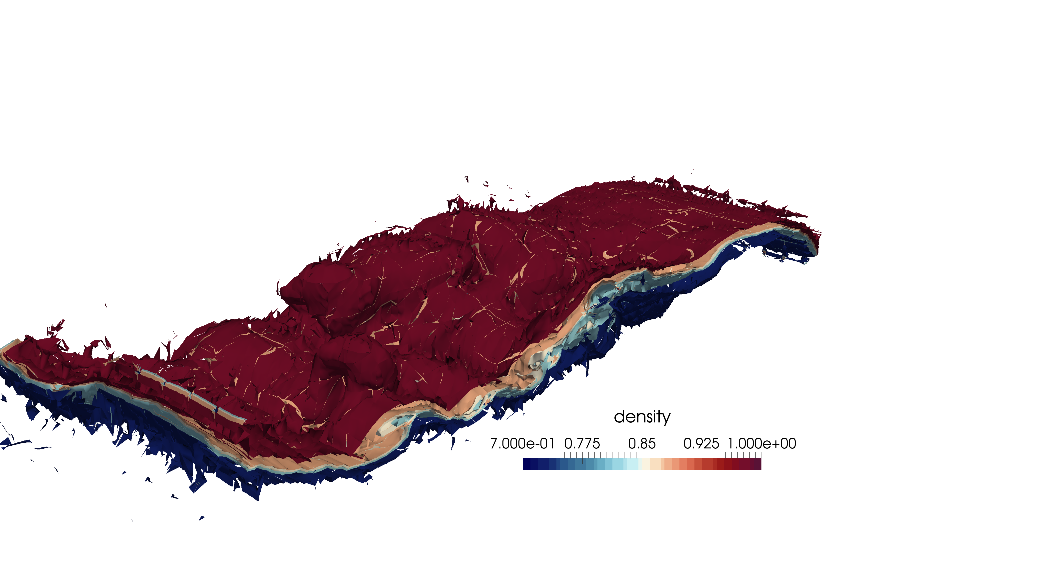}}    
\end{subfigure}

\begin{subfigure}[]{
      \label{fig:q_iso_6000_nomod}
      \includegraphics[trim=0cm 0cm 4cm 2cm, clip=true, totalheight=0.35\textheight]{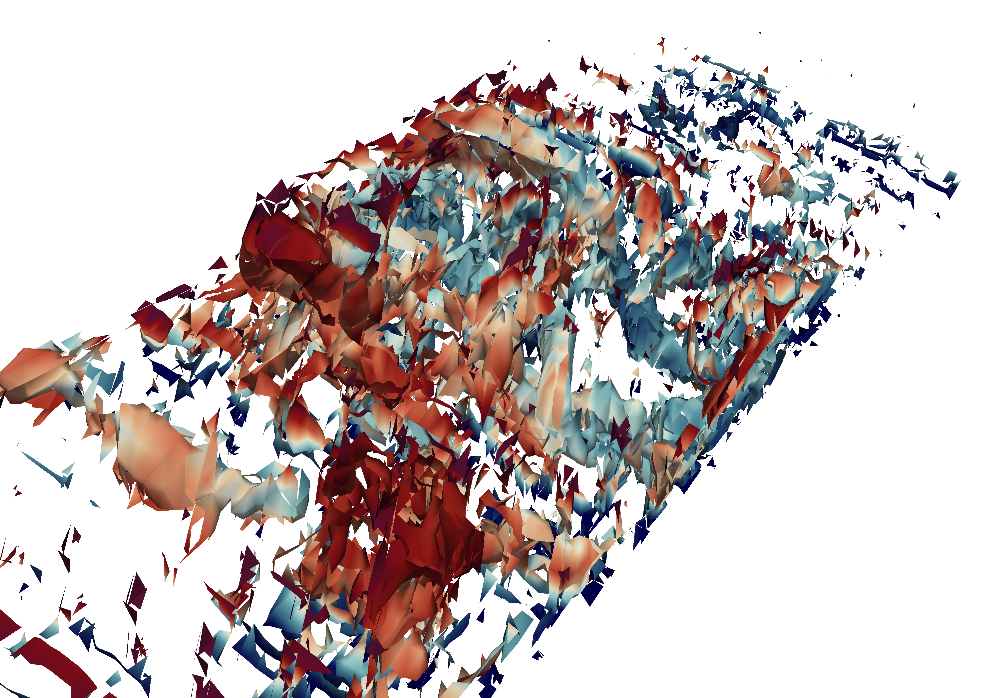}}
\end{subfigure} 
\caption{Isosurfaces at $t=4$ for $Re=6000$ (No model LES) (a) density isosurfaces ($\rho = 0.72, 0.82,0.85, 0.88, 0.98$). (b) $Q=20$ isosurface (colored by density).}
\end{figure}

Considering the $Re=6000$ case at $t=4,$
if we look at the density profile obtained with the Smagorinsky model (figure \ref{fig:density_iso_6000_smag}), we can notice that it is much smoother with respect to the DNS one, suggesting the fact that the Smagorinsky model could be too dissipative also in this case. This seems to be  confirmed by the  behaviour of the $Q=20$ isosurface.
If we compare the Smagorinsky one (figure \ref{fig:q_iso_6000_smag}) with that of the DNS (figure \ref{fig:q_iso_6000_dns}), we can see that the Smagorinsky model yields less turbulent structures with respect to the DNS.
An improvement can be obtained employing dynamic models, both in the density and  $Q$ fields. In particular, figure \ref{fig:q_iso_6000_dyn} shows  that the dynamic isotropic model provides many more turbulent structures with respect to the Smagorinsky model and allows the representation of the majority of the turbulent structures actually present in the DNS, even if the structures in the $Q$ field appear more scattered. 
The same considerations can be made for the anisotropic dynamic model (see figures \ref{fig:density_iso_6000_aniso} and \ref{fig:q_iso_6000_aniso} for the density and $Q=20$ isosurface respectively), even if a slightly more dissipative character is present, especially in the $Q$ profile, with respect to the isotropic version.
Notice that, even though the introduction of a dynamic model provides better results with respect to the Smagorinsky model, if we look at the profiles of density and $Q$ obtained with the no-model LES (figures \ref{fig:density_iso_6000_nomod} and \ref{fig:q_iso_6000_nomod}) and we compare them with those obtained with the dynamic models, it is not completely clear, from these instantaneous fields, if the introduction of a dynamic model leads to some improvement with respect to a simple no-model LES.

\begin{figure}
\centering
\begin{subfigure}[]{
     \label{fig:density_iso_6000_smag_9}
     \includegraphics[trim=1.8cm 1cm 2cm 1cm, clip=true, totalheight=0.42\textheight]{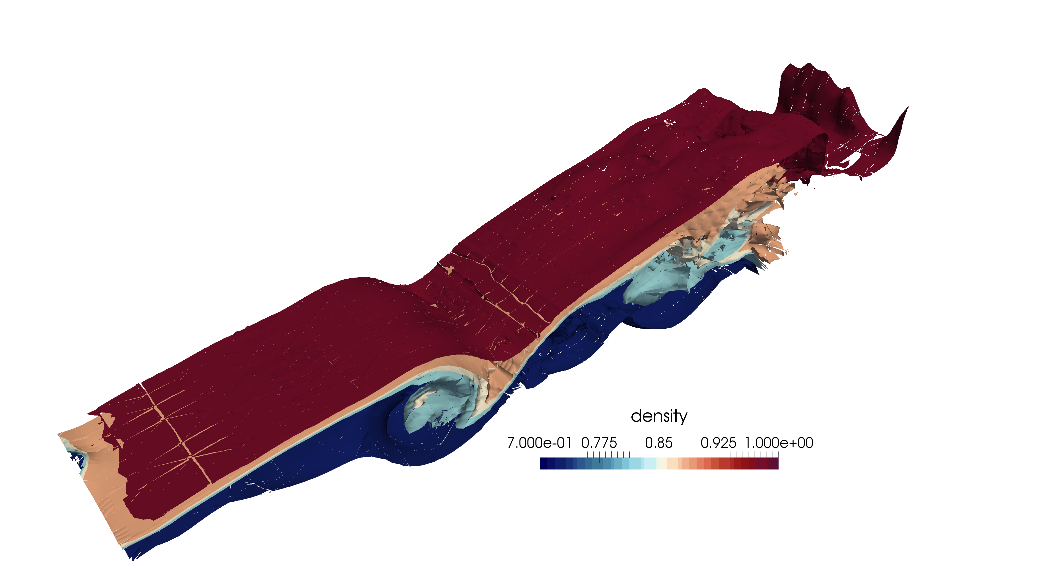}}    
\end{subfigure}

\begin{subfigure}[]{
      \label{fig:q_iso_6000_smag_9}
      \includegraphics[trim=2.6cm 1cm 2cm 1.4cm, clip=true, totalheight=0.42\textheight]{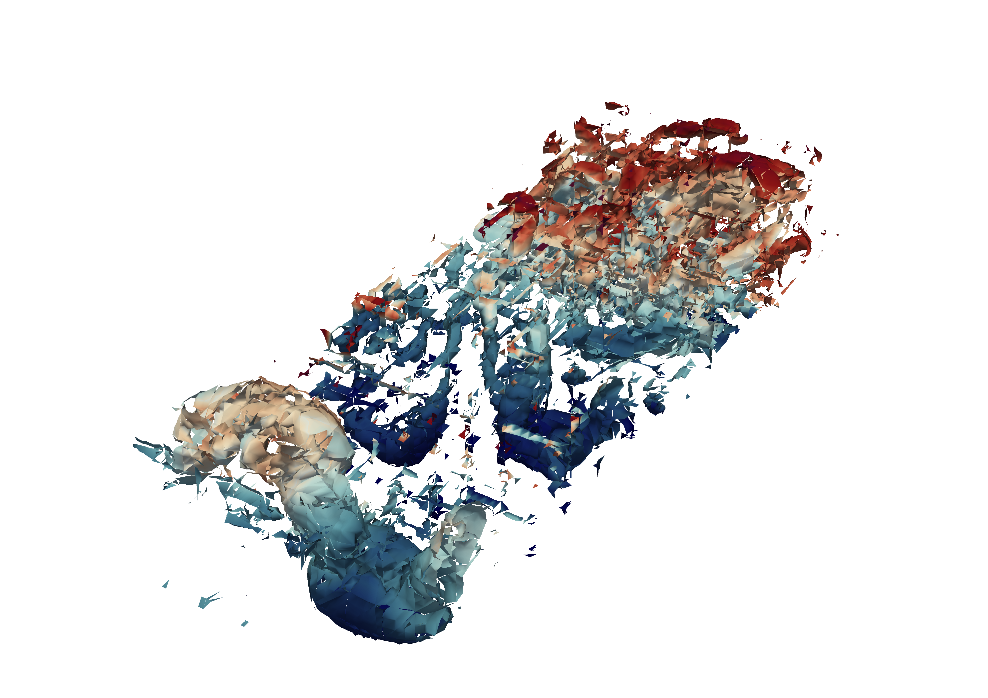}}
\end{subfigure} 
\caption{Isosurfaces at $t=9$ for $Re=6000$ (Smagorinsky model LES) (a) density isosurfaces ($\rho = 0.72, 0.82,0.85, 0.88, 0.98$). (b) $Q=5$ isosurface (colored by density).}
\end{figure}

\begin{figure}
\centering
\begin{subfigure}[]{
     \label{fig:density_iso_6000_dyn_9}
     \includegraphics[trim=1.8cm 1cm 2cm 1cm, clip=true, totalheight=0.42\textheight]{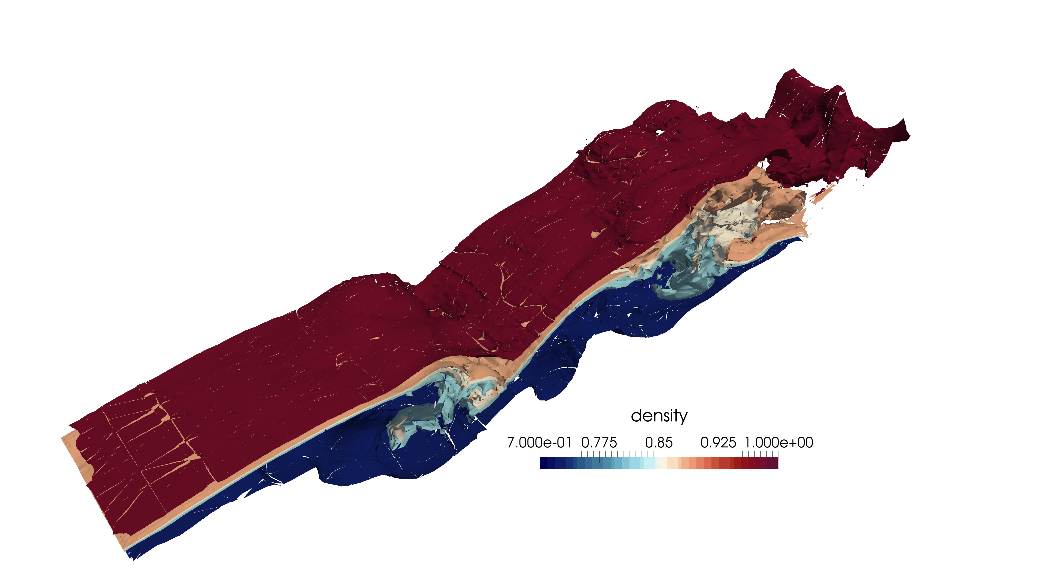}}    
\end{subfigure}

\begin{subfigure}[]{
      \label{fig:q_iso_6000_dyn_9}
      \includegraphics[trim=2.6cm 1cm 2cm 1.4cm, clip=true, totalheight=0.42\textheight]{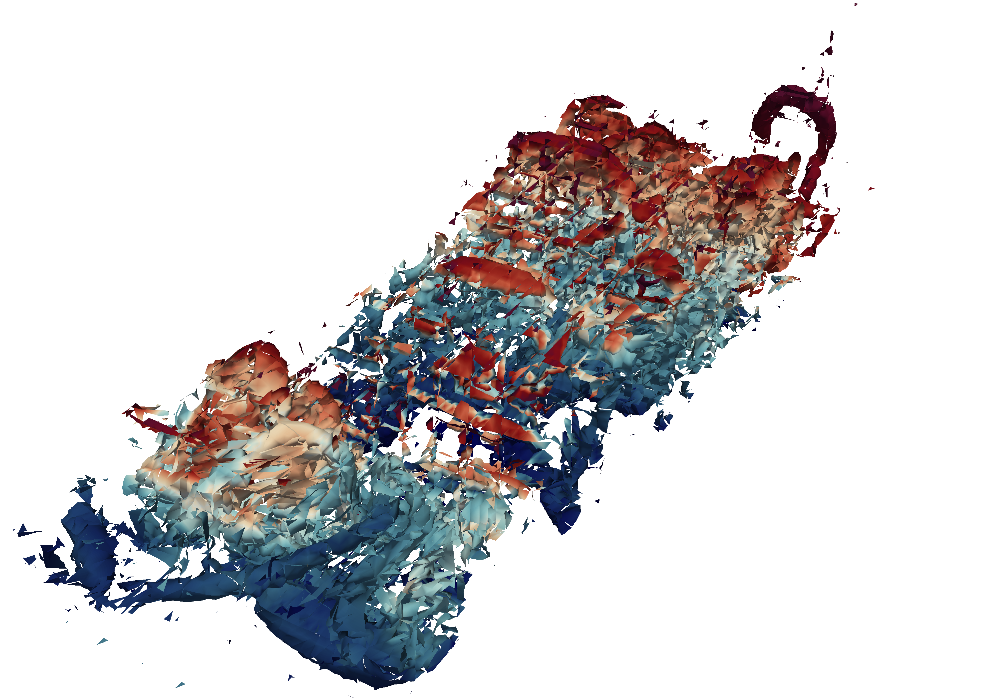}}
\end{subfigure} 
\caption{Isosurfaces at $t=9$ for $Re=6000$ (Isotropic dynamic model LES) (a) density isosurfaces ($\rho = 0.72, 0.82,0.85, 0.88, 0.98$). (b) $Q=5$ isosurface (colored by density).}
\end{figure}

\begin{figure}
\centering
\begin{subfigure}[]{
     \label{fig:density_iso_6000_aniso_9}
     \includegraphics[trim=1.8cm 1cm 2cm 1cm, clip=true, totalheight=0.42\textheight]{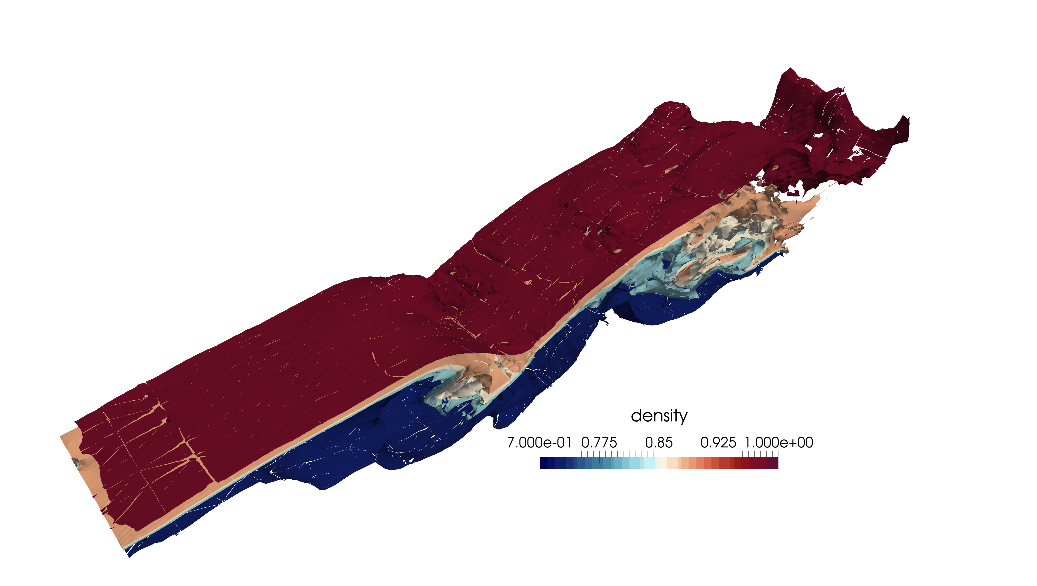}}    
\end{subfigure}

\begin{subfigure}[]{
      \label{fig:q_iso_6000_aniso_9}
      \includegraphics[trim=2.6cm 1cm 2cm 1.4cm, clip=true, totalheight=0.42\textheight]{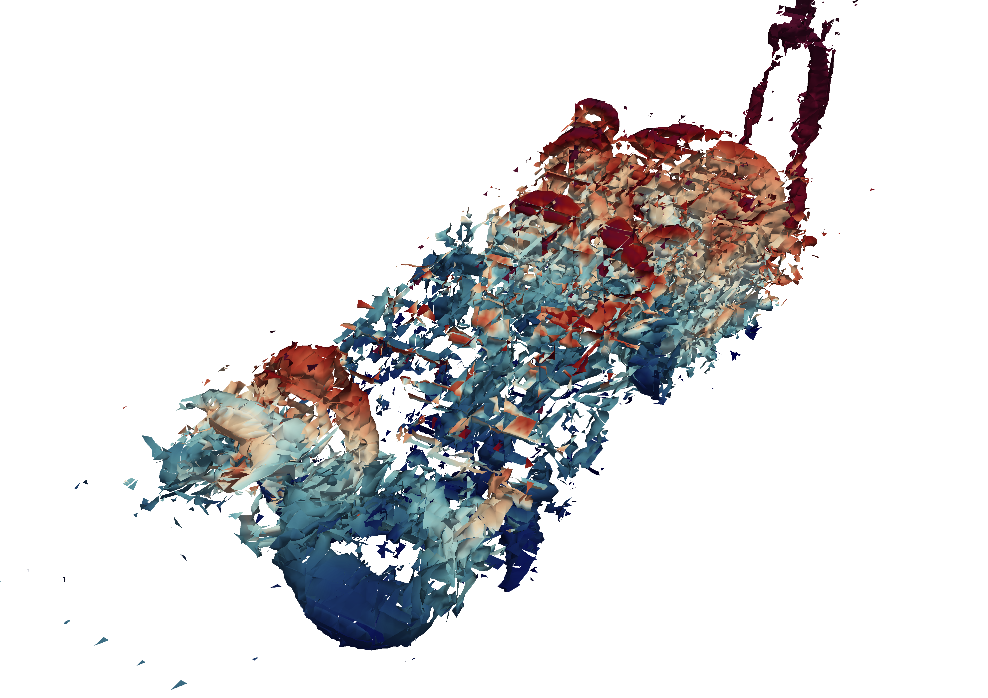}}
\end{subfigure} 
\caption{Isosurfaces at $t=9$ for $Re=6000$ (Anisotropic dynamic model LES) (a) density isosurfaces ($\rho = 0.72, 0.82,0.85, 0.88, 0.98$). (b) $Q=5$ isosurface (colored by density).}
\end{figure}

\begin{figure}
\centering
\begin{subfigure}[]{
     \label{fig:density_iso_6000_nomod_9}
     \includegraphics[trim=1.8cm 1cm 2cm 1cm, clip=true, totalheight=0.42\textheight]{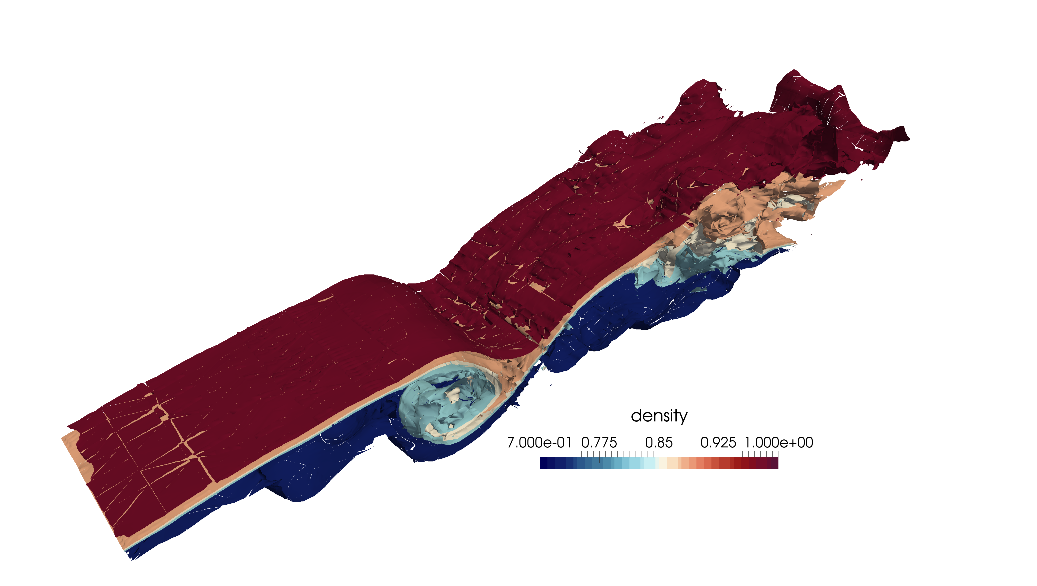}}    
\end{subfigure}

\begin{subfigure}[]{
      \label{fig:q_iso_6000_nomod_9}
      \includegraphics[trim=2.6cm 1cm 2cm 1.4cm, clip=true, totalheight=0.42\textheight]{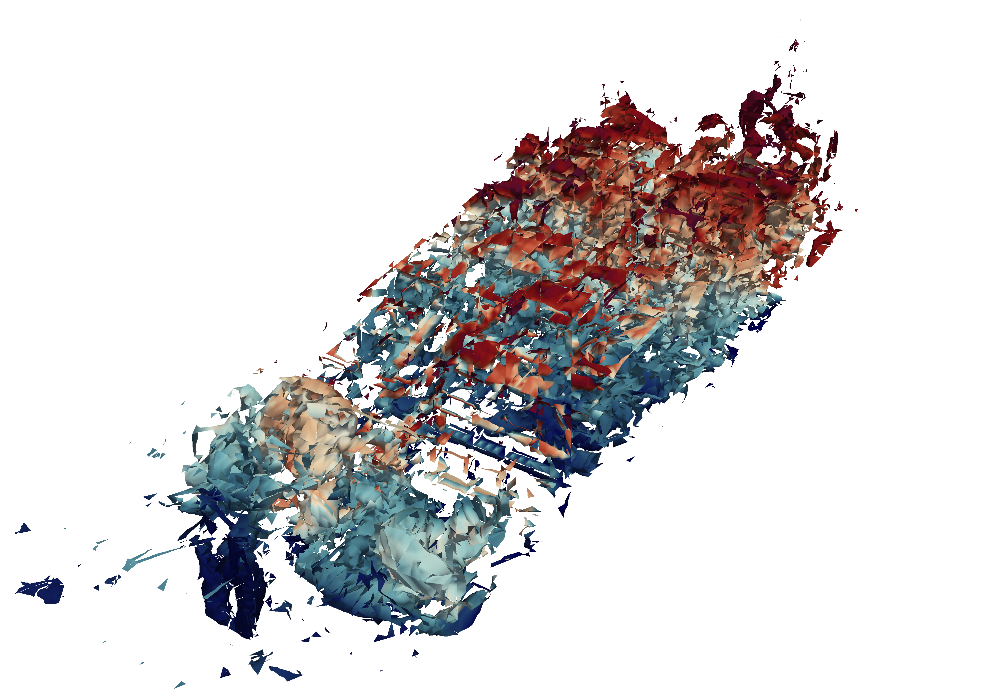}}
\end{subfigure} 
\caption{Isosurfaces at $t=9$ for $Re=6000$ (No model LES) (a) density isosurfaces ($\rho = 0.72, 0.82,0.85, 0.88, 0.98$). (b) $Q=5$ isosurface (colored by density).}
\end{figure}

Considering then the $Re=6000$ case at $t=9,$
if we compare figures \ref{fig:q_iso_6000_smag_9}, which represents the $Q=5$ isosurface obtained with the Smagorinsky model, with the corresponding DNS field (figure \ref{fig:q_iso_6000_dns_9}), we can confirm the fact that the Smagorinsky model is too dissipative. 
As for $t=4$, an improvement can be obtained employing a dynamic model, both in terms of density and $Q$ (see figures \ref{fig:q_iso_6000_dyn_9} and \ref{fig:q_iso_6000_aniso_9}). Notice that, at $t=9$, some differences arise also between the dynamic models results and the no-model LES results. In particular, if we compare the $Q=5$ field obtained with the isotropic dynamic model (figure \ref{fig:q_iso_6000_dyn_9}) and the anisotropic dynamic model (figure \ref{fig:q_iso_6000_aniso_9}) with the corresponding no-model field (figure \ref{fig:q_iso_6000_nomod_9}), we can see that the introduction of a model leads to the presence of more coherent turbulent structures with respect to the no-model case. However,it appears difficult to conclude,  based only on  analysis of the instantaneous fields,  whether the introduction of a dynamic model actually leads to better results with respect to the no-model LES.

\begin{figure}
\centering
\includegraphics[trim=2.6cm 0.6cm 2cm 0cm, clip=true, totalheight=0.48\textheight]{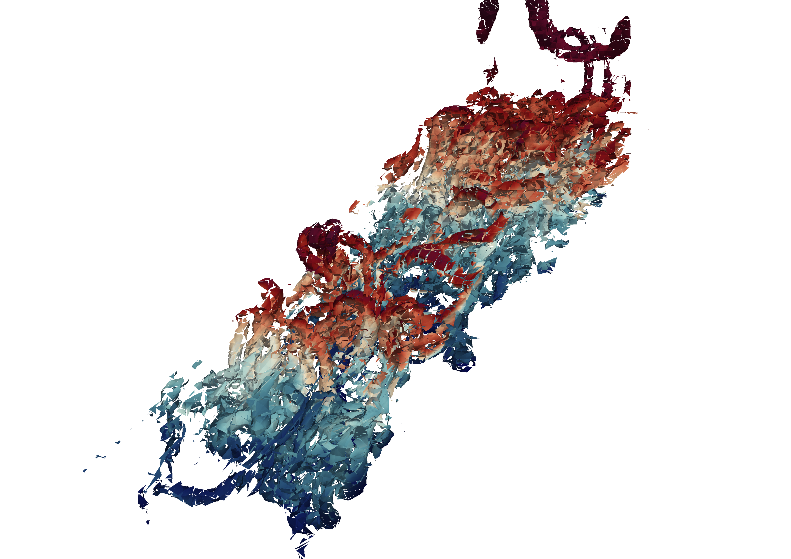}
\caption{Post-filtered DNS isosurfaces of $Q=5$ at $t=9$ for $Re=6000$.}
\label{fig:postfilter}
\end{figure}

In order to allow for a fairer comparison of  the LES results, we have also filtered the DNS results
{\it a posteriori}, to see if additional informations on the performances of the different turbulence models can be obtained with respect to the comparison with the simple DNS $Q$ field. Since it would have been quite cumbersome to post filter the DNS results considering the same polynomial degree ($p=4$) and a coarser grid, we simply project the solution onto a lower dimensional polynomial space, in order to obtain a number of DOFs similar to the one of the $Re=6000$ LES. In figure \ref{fig:postfilter} we present the post-filtered DNS field for $Re=6000$ and $t=9$. If we compare this field to the corresponding DNS field (figure\ref{fig:q_iso_6000_dns_9}), we can notice that some small scale structures are less evident, as expected, even though the larger scale structure can still be identifed. However if we compare the DNS post filtered field of figure \ref{fig:postfilter} with the corresponding LES fields of figures \ref{fig:q_iso_6000_smag_9}, \ref{fig:q_iso_6000_dyn_9}, \ref{fig:q_iso_6000_aniso_9}, \ref{fig:q_iso_6000_nomod_9}, we cannot reach a satisfying conclusion about the performances of the different models.

\begin{figure}[]
\centering
\begin{subfigure}[]{
      \label{fig:dissipated_energy_3d_3000}
      \includegraphics[width=0.9\textwidth]{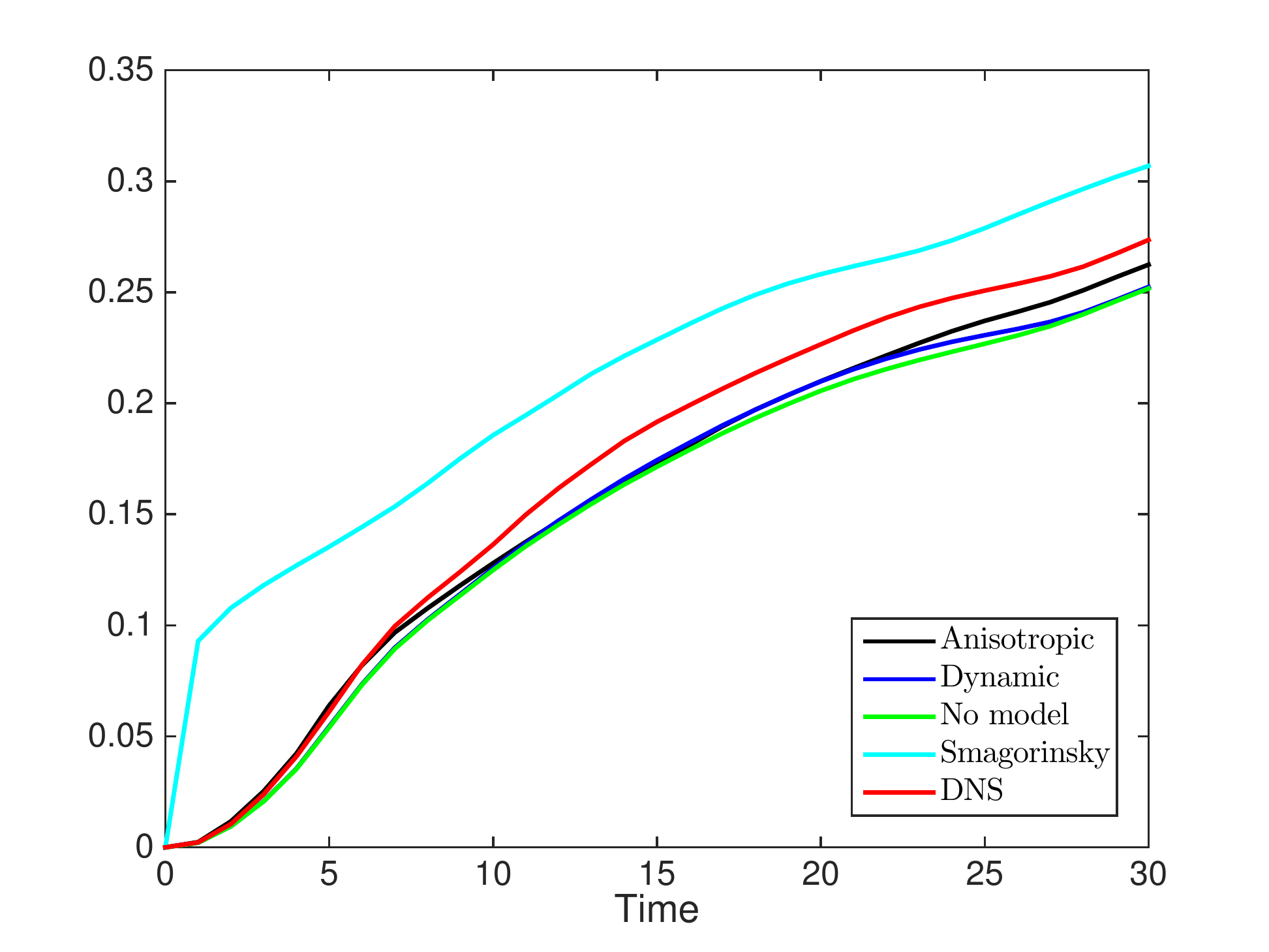}}
\end{subfigure} 
\begin{subfigure}[]{
        \label{fig:dissipated_energy_3d_6000}
       \includegraphics[width=0.9\textwidth]{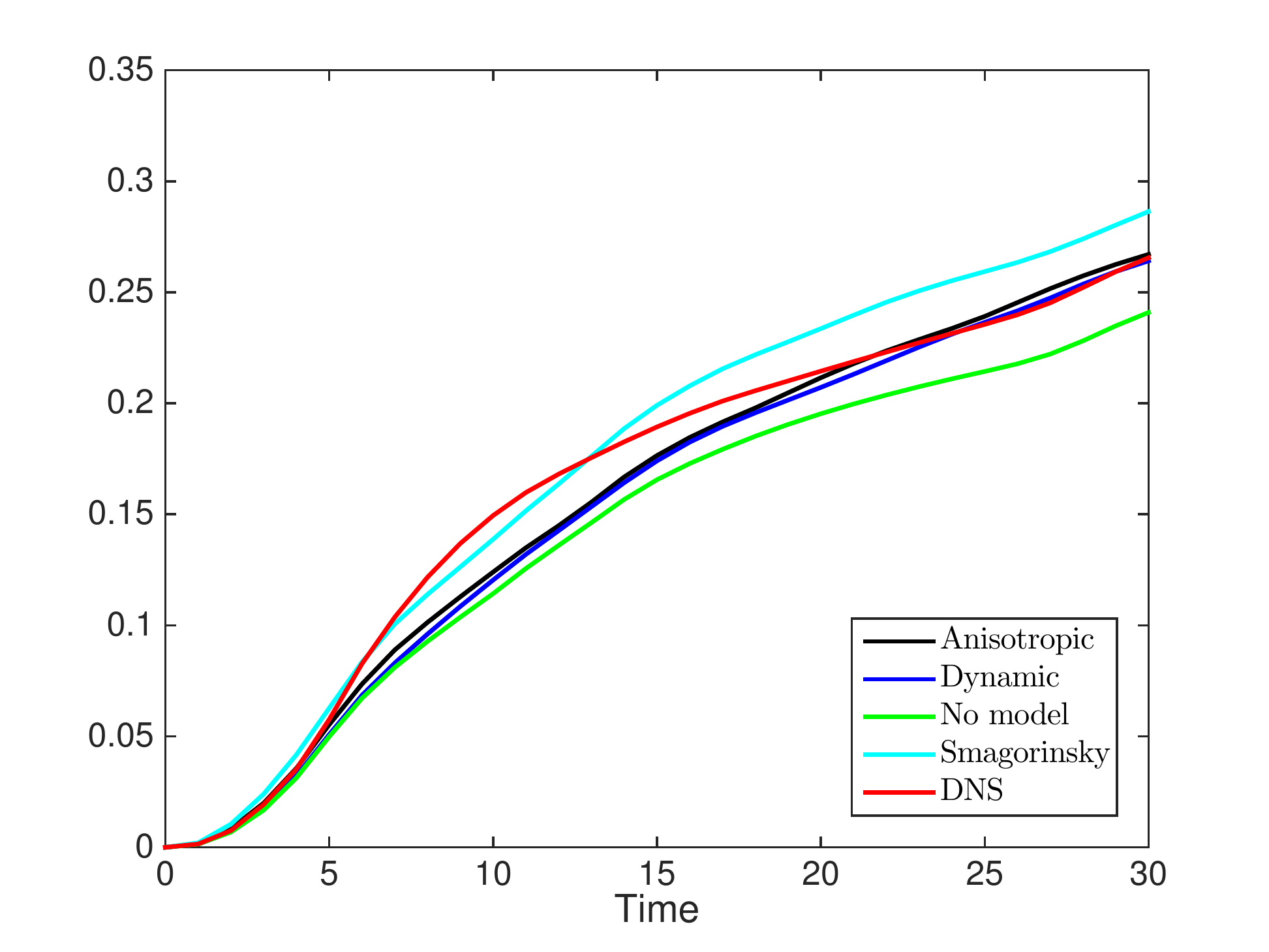}}
\end{subfigure}
\caption{Dissipated energy as a function of time. (a) $Re=3000$. (b) $Re=6000$.}
\label{fig:dissipated_energy_3d}
\end{figure}

In order to try to better discriminate the behaviour of the different LES models and their performances with respect to the DNS results, we consider the dissipated energy profiles as a function of time, computed as in equation (\ref{eq:dissipated_energy_les}) for the no-model case, the Smagorinsky model and the isotropic dynamic model, and as in equation (\ref{eq:dissipated_energy_aniso}) for the anisotropic dynamic model. The dissipated energy profiles are presented both for $Re=3000$ and $Re=6000$ cases in figure \ref{fig:dissipated_energy_3d}. 

Concerning the $Re=3000$ case (figure \ref{fig:dissipated_energy_3d_3000}), we can confirm the finding that the Smagorinsky model is far too dissipative: the dissipated energy obtained with the Smagorinsky model (cyan curve) is always greater than the DNS one (red curve). 
Considering the isotropic dynamic model (blue curve), we can notice that its dissipated energy profile is quite similar to the one provided by the no-model LES (green curve). Even if slightly more dissipative, the isotropic dynamic model is not able to recover the significant increase in the dissipated energy of the DNS starting from $t \simeq 7$. 
The anisotropic dynamic model (black curve) seems the one that provides the better results in terms of dissipated energy: the initial increasing in the DNS dissipated energy is quite well reproduced even though also the anisotropic dynamic model underestimates the energy dissipated in the DNS starting from $t \simeq 7$. The anisotropic dynamic model curve seems the one closer to the DNS profile also for $t>22$.

Considering the $Re=6000$ simulations (figure \ref{fig:dissipated_energy_3d_6000}), the behaviour of the Smagorinsky model (cyan curve) is always too dissipative, especially if we consider times $t > 14$. 
The no-model LES (green curve) and the LES obtained with the isotropic dynamic model (blue curve) present results with larger differences between each other than in the $Re=3000$ case. The isotropic dynamic model presents a more dissipative behaviour with respect to the no model case. 
Both the isotropic and the anisotropic dynamic model (black curve) appear to provide the best results with a good reproduction of the initial ($t<5$) and final ($t>20$) DNS dissipated energy profile, even if they are not able to capture the behaviour of the DNS dissipated energy for $t \in [5,20]$.
 
\begin{figure}[]
\centering
\begin{subfigure}[]{
       \label{fig:max_nuratio-3000}
      \includegraphics[width=0.9\textwidth]{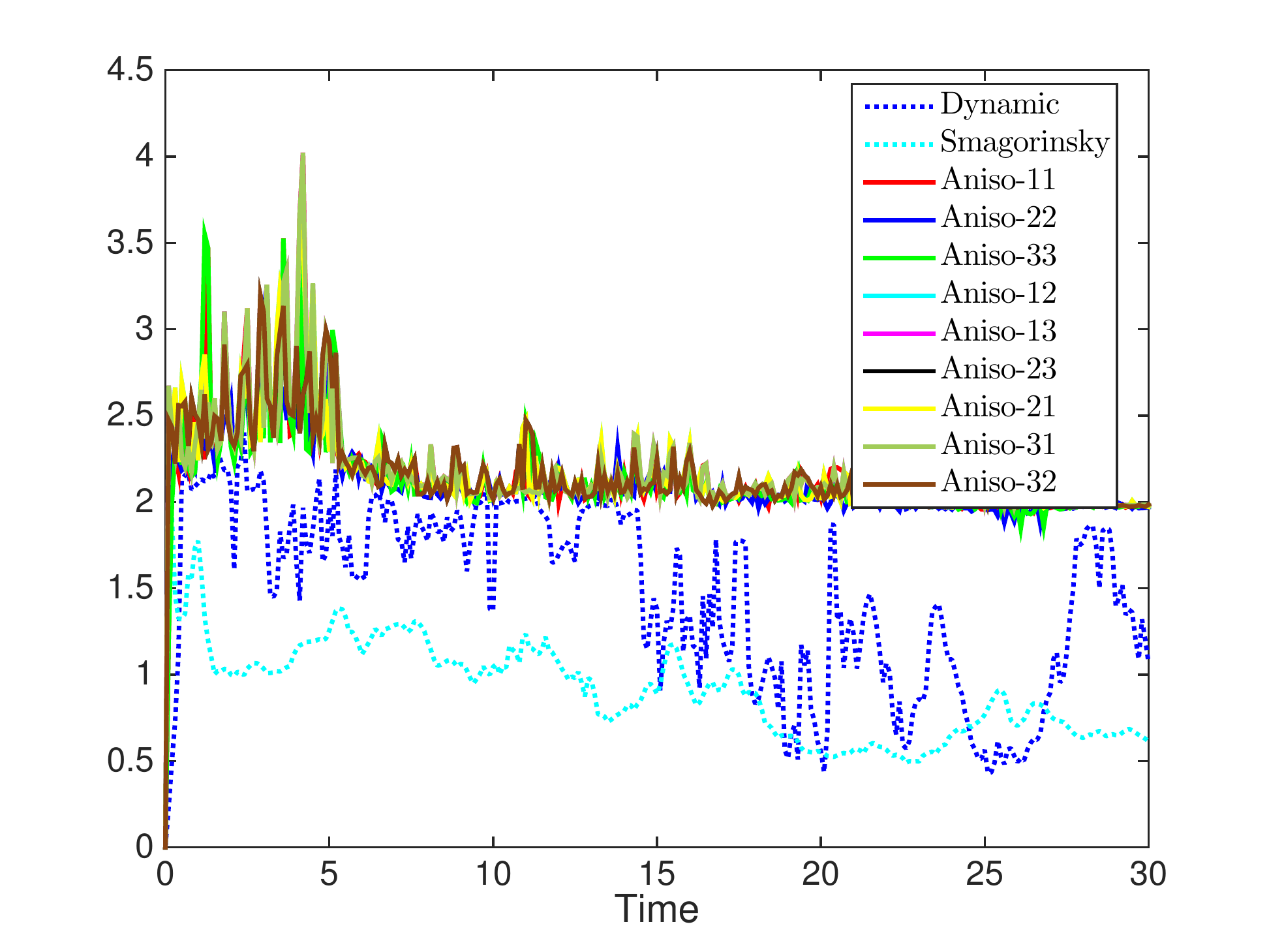}}
\end{subfigure}
\begin{subfigure}[]{
      \label{fig:max_nuratio-6000}
      \includegraphics[width=0.9\textwidth]{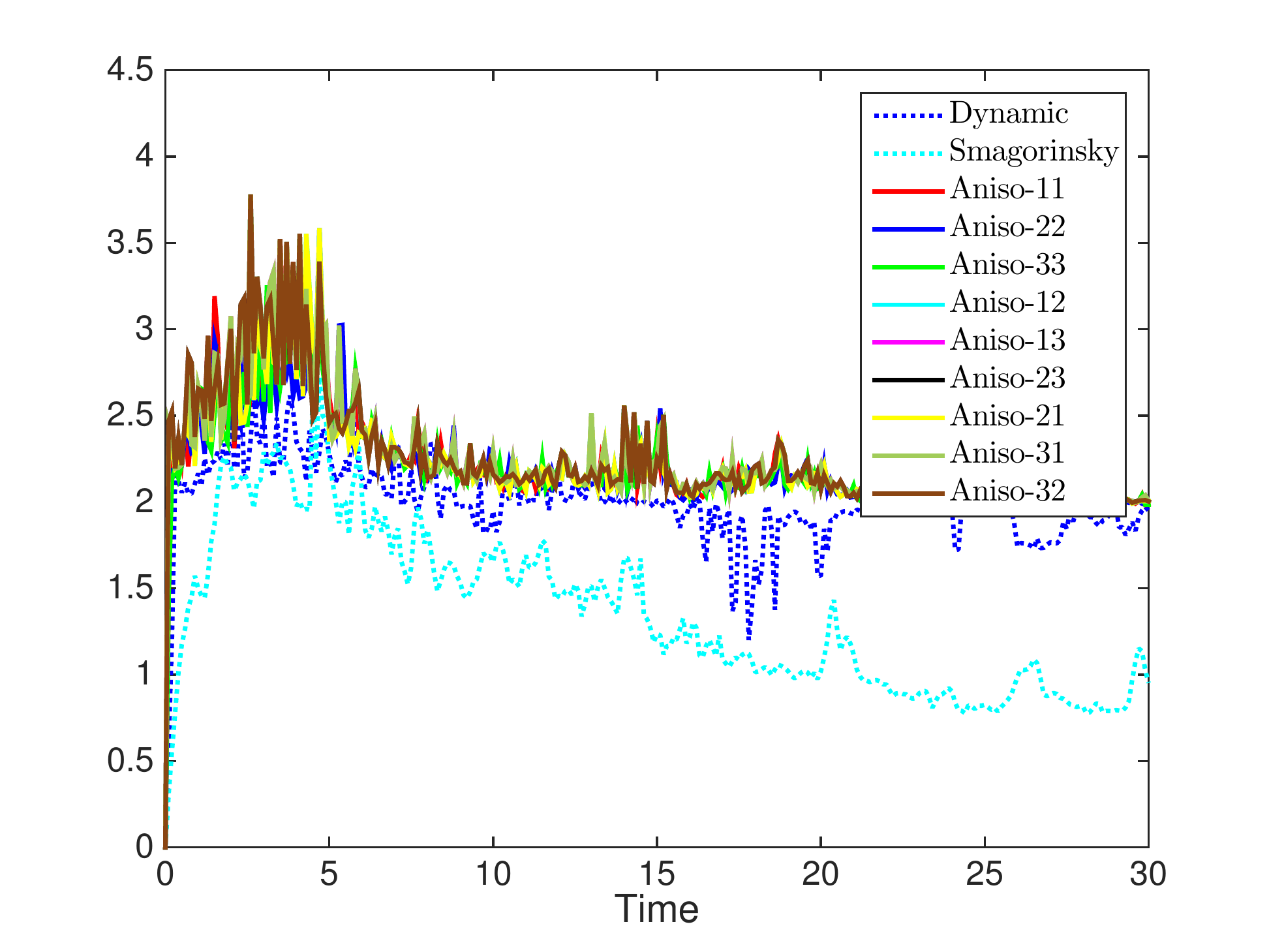}}
\end{subfigure} 
\caption{Maximum value over the whole domain of $\nu_{\rm sgs}/\nu$, as a function of time. (a) $Re=3000$. (b) $Re=6000$.}
\label{fig:max_nuratio}
\end{figure}

As an additional diagnostic, we have  also computed the temporal evolution of the maximum and minimum values of the ratio $\nu_{\rm sgs}/\nu$ between the subgrid-scale viscosity $\nu_{sgs}$ and the molecular viscosity $\nu$ over the whole domain (see figure \ref{fig:max_nuratio}).
Figure \ref{fig:max_nuratio-3000} corresponds to the $Re=3000$ case, while figure \ref{fig:max_nuratio-6000} to Reynolds $Re=6000$. Solid lines correspond to the different viscosities related to the anisotropic model (for a three-dimensional problem we have six different components of the tensor of the subgrid viscosities, see appendix \ref{section:aniso_model}), the blue dotted line corresponds to the isotropic dynamic model, while the cyan dotted curve refers to the Smagorinsky model.  
The anisotropic dynamic model provides the larger peak values, the isotropic dynamic model intermediate peak values while the Smagorinsky model the lowest values. We can notice that 
greater differences are present between the peak values of the isotropic and the ansotropic dynamic models in the Reynolds $Re=3000$ case (figure \ref{fig:max_nuratio-3000}), with respect to the $Re=6000$ case. This probably means that, in the $Re=6000$ case (figure \ref{fig:max_nuratio-6000}), the flow has a more isotropic character than in the $Re=3000$ case.
Notice that, for both the $Re=3000$ and the $Re=6000$ case and especially for the anisotropic dynamic model, the maximum values of the ratio $\nu_{\rm sgs}/\nu$ tend to a constant value toward the end of the simulation. This trend is evident also for the Smagorinsky and the dynamic isotropic models even if larger amplitude oscillations are present.

\begin{figure}[]
\centering
\begin{subfigure}[]{
       \label{fig:min_nuratio-3000}
      \includegraphics[width=0.9\textwidth]{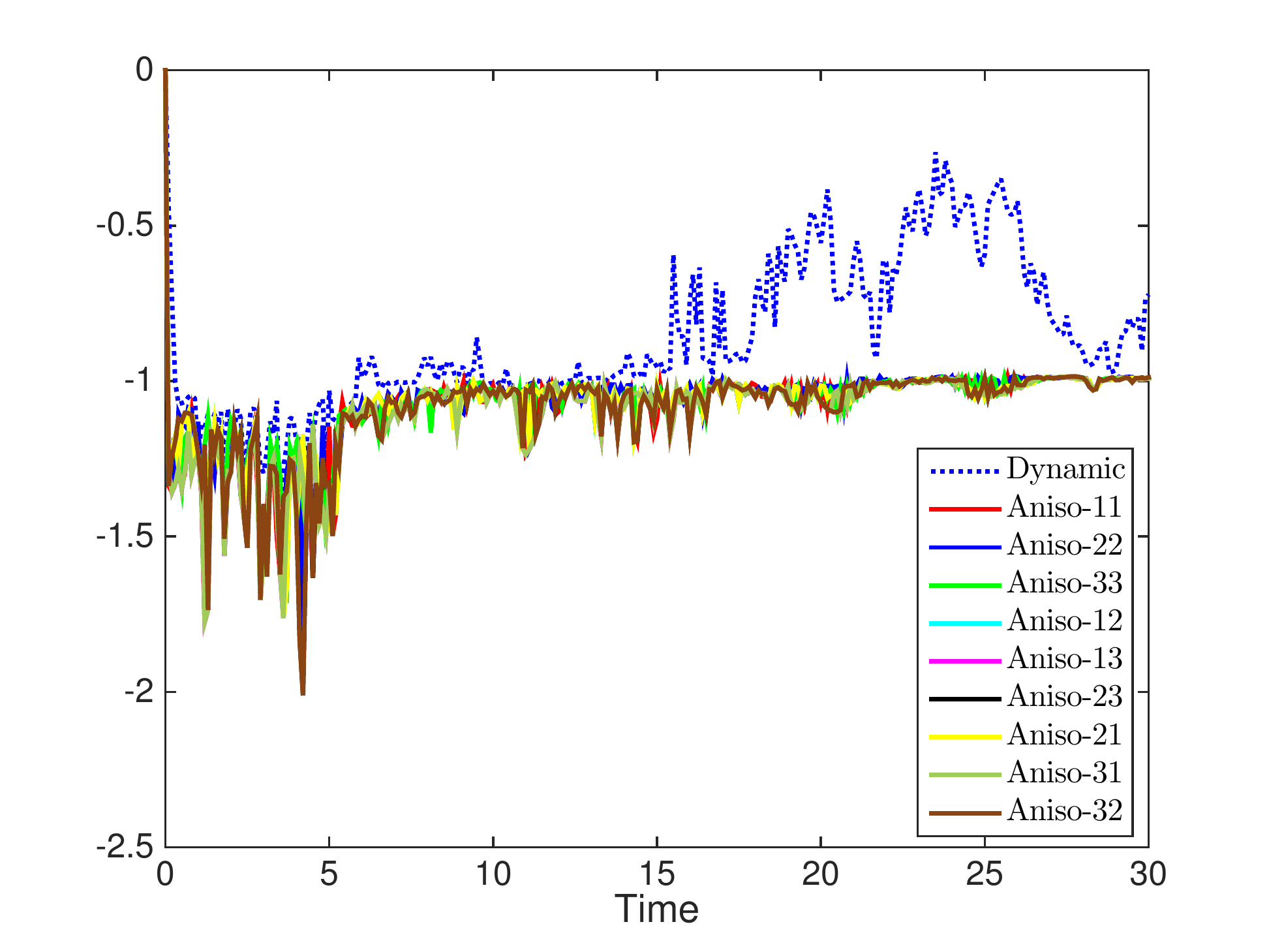}}
\end{subfigure}
\begin{subfigure}[]{
      \label{fig:min_nuratio-6000}
      \includegraphics[width=0.9\textwidth]{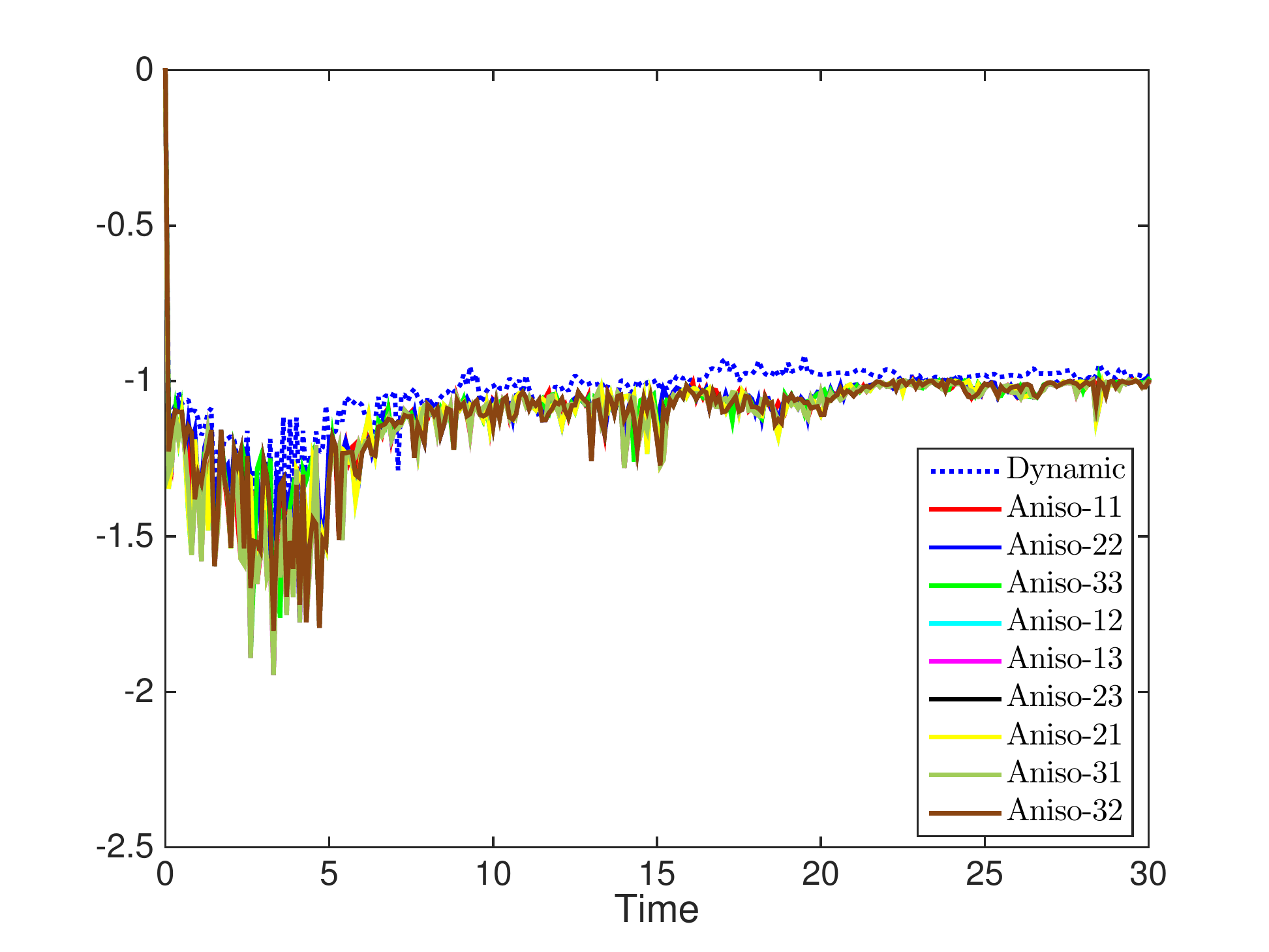}}
\end{subfigure} 
\caption{Minimum value value over the whole domain of $\nu_{\rm sgs}/\nu$, as a function of time. (a) $Re=3000$. (b) $Re=6000$.}
\label{fig:min_nuratio}
\end{figure}

\begin{figure}[]
\centering
\begin{subfigure}[]{
        \label{fig:nuratio_a}
       \includegraphics[trim=7cm 20cm 1cm 20cm, clip=true, totalheight=0.17\textheight]{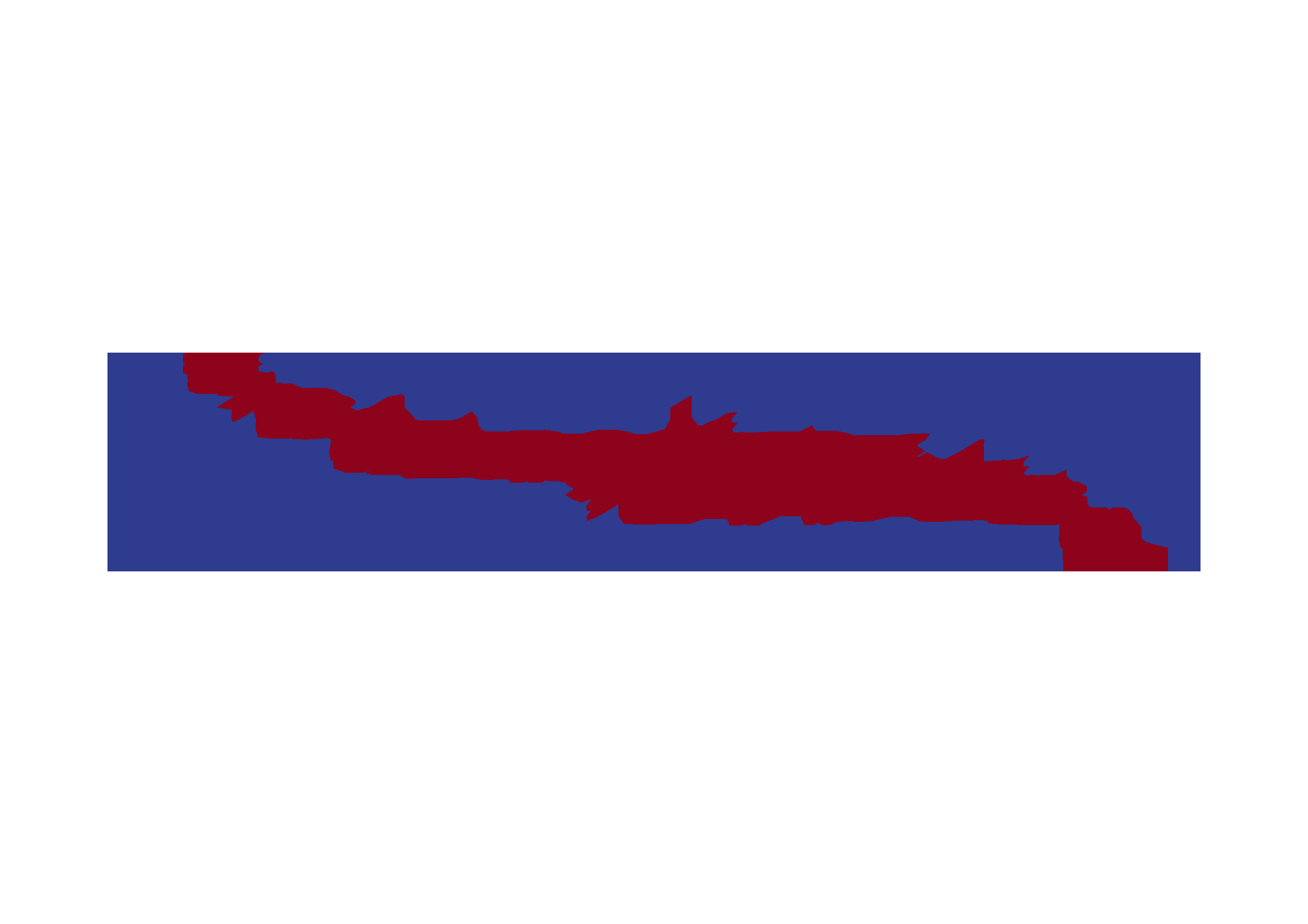}}
\end{subfigure}
\begin{subfigure}[]{
        \label{fig:nuratio_b}
       \includegraphics[trim=7cm 20cm 1cm 20cm, clip=true, totalheight=0.17\textheight]{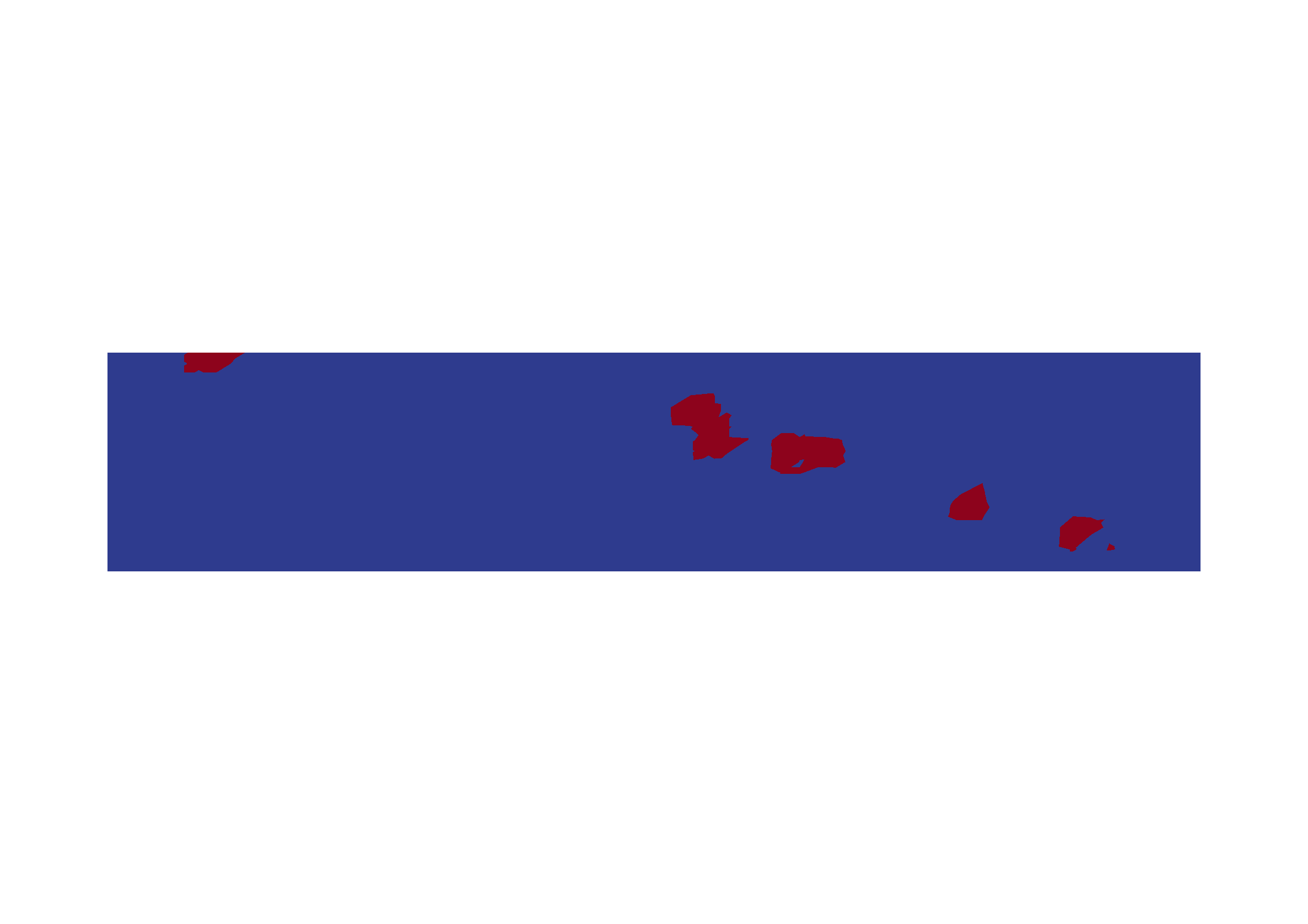}}
\end{subfigure}
\begin{subfigure}[]{
      \label{fig:nuratio_c}
      \includegraphics[trim=7cm 20cm 1cm 20cm, clip=true, totalheight=0.17\textheight]{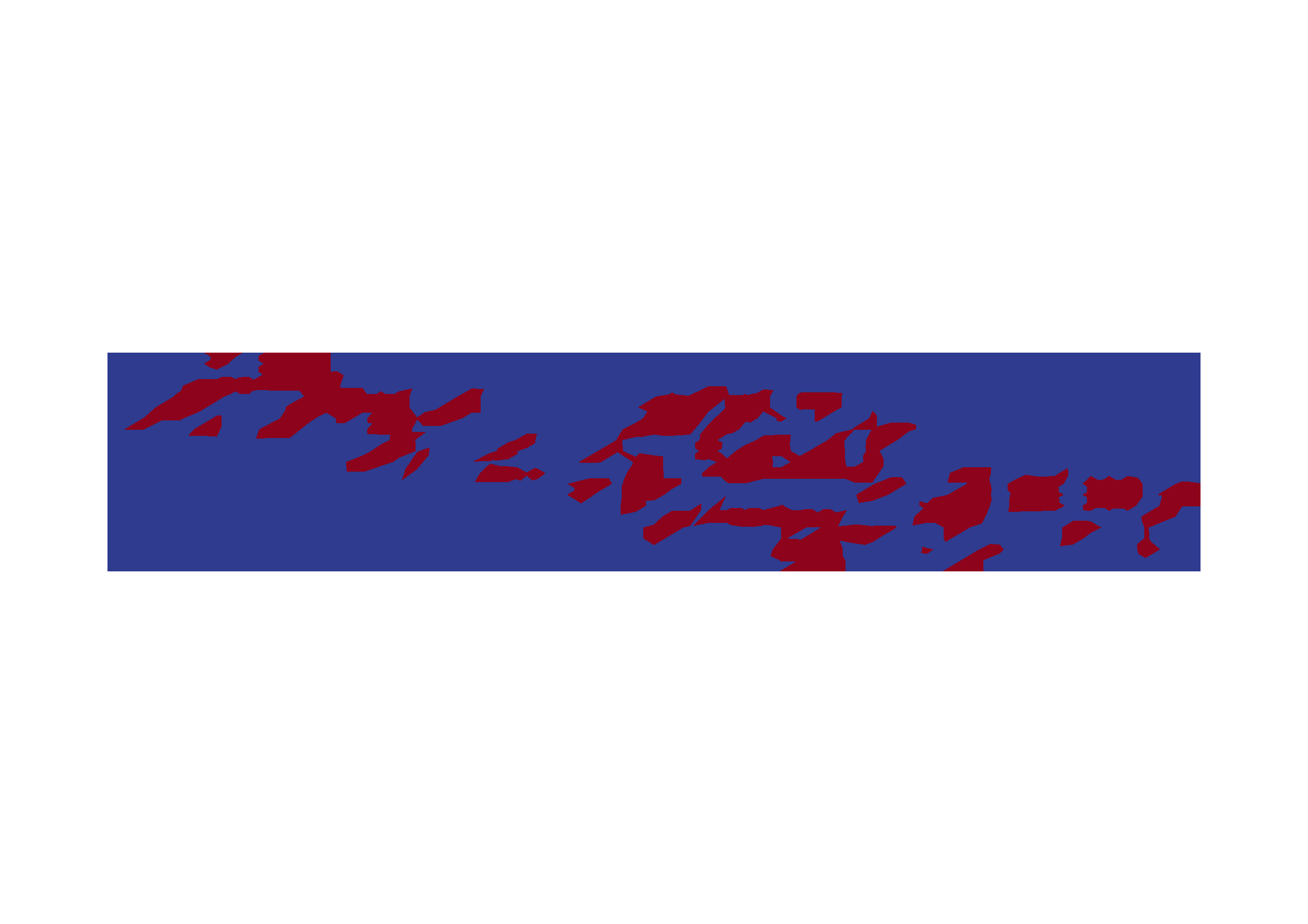}}
\end{subfigure}
\caption{Portion of the domain (in red), on the plane $y=0.5$, in which the condition $\nu_{sgs}/\nu > 0.3$ is satisfied for $Re=3000$ at $t=4$. (a) Smagorinsky model. (b) Isotropic dynamic model. (c) Anisotropic dynamic model (component $22$ of $\nu_{sgs}/\nu$).}
\label{fig:nuratio}
\end{figure}

\begin{figure}[]
\centering
\begin{subfigure}[]{
        \label{fig:nuratio_a_neg}
       \includegraphics[trim=7cm 20cm 1cm 20cm, clip=true, totalheight=0.17\textheight]{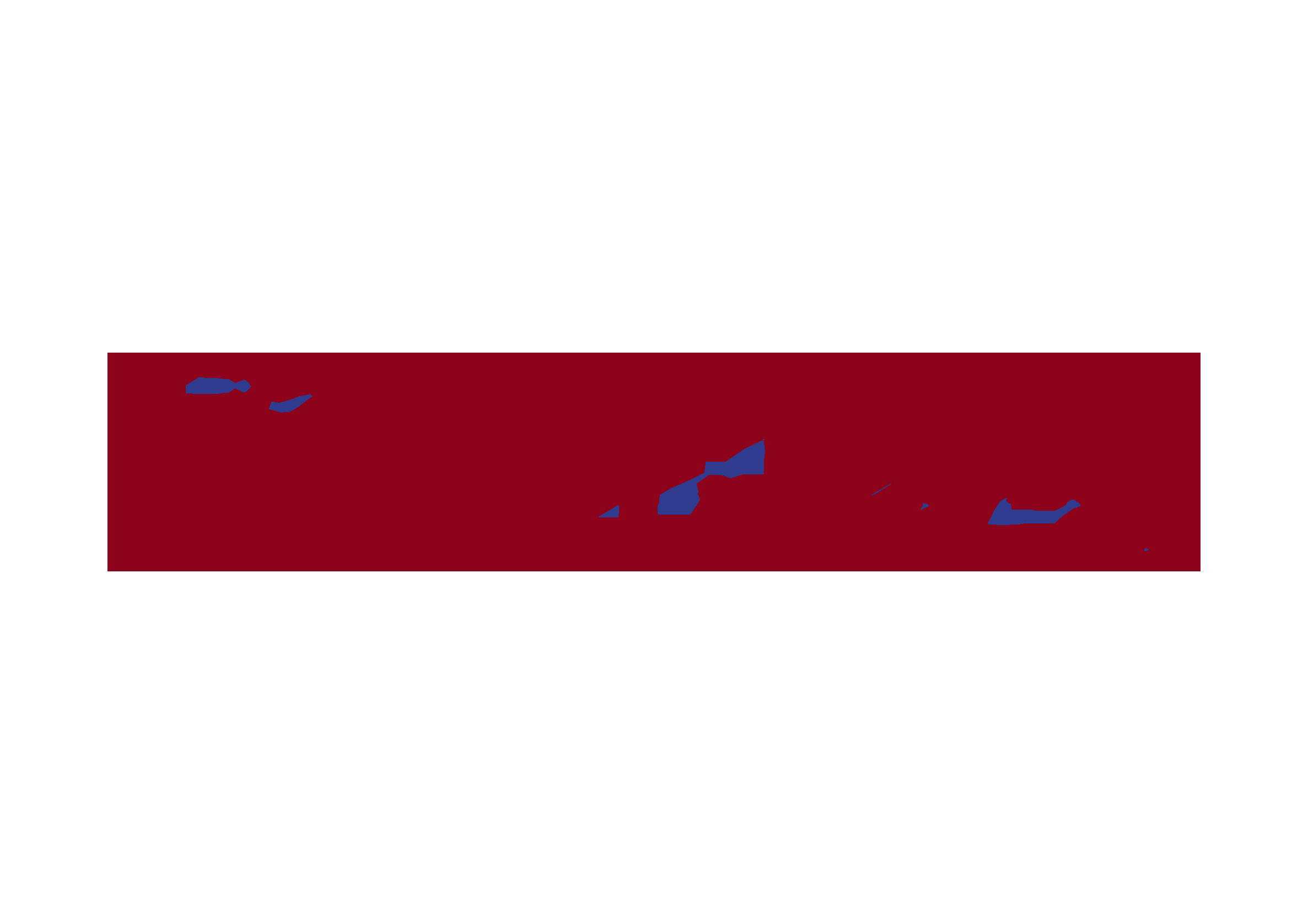}}
\end{subfigure}
\begin{subfigure}[]{
        \label{fig:nuratio_b_neg}
       \includegraphics[trim=7cm 20cm 1cm 20cm, clip=true, totalheight=0.17\textheight]{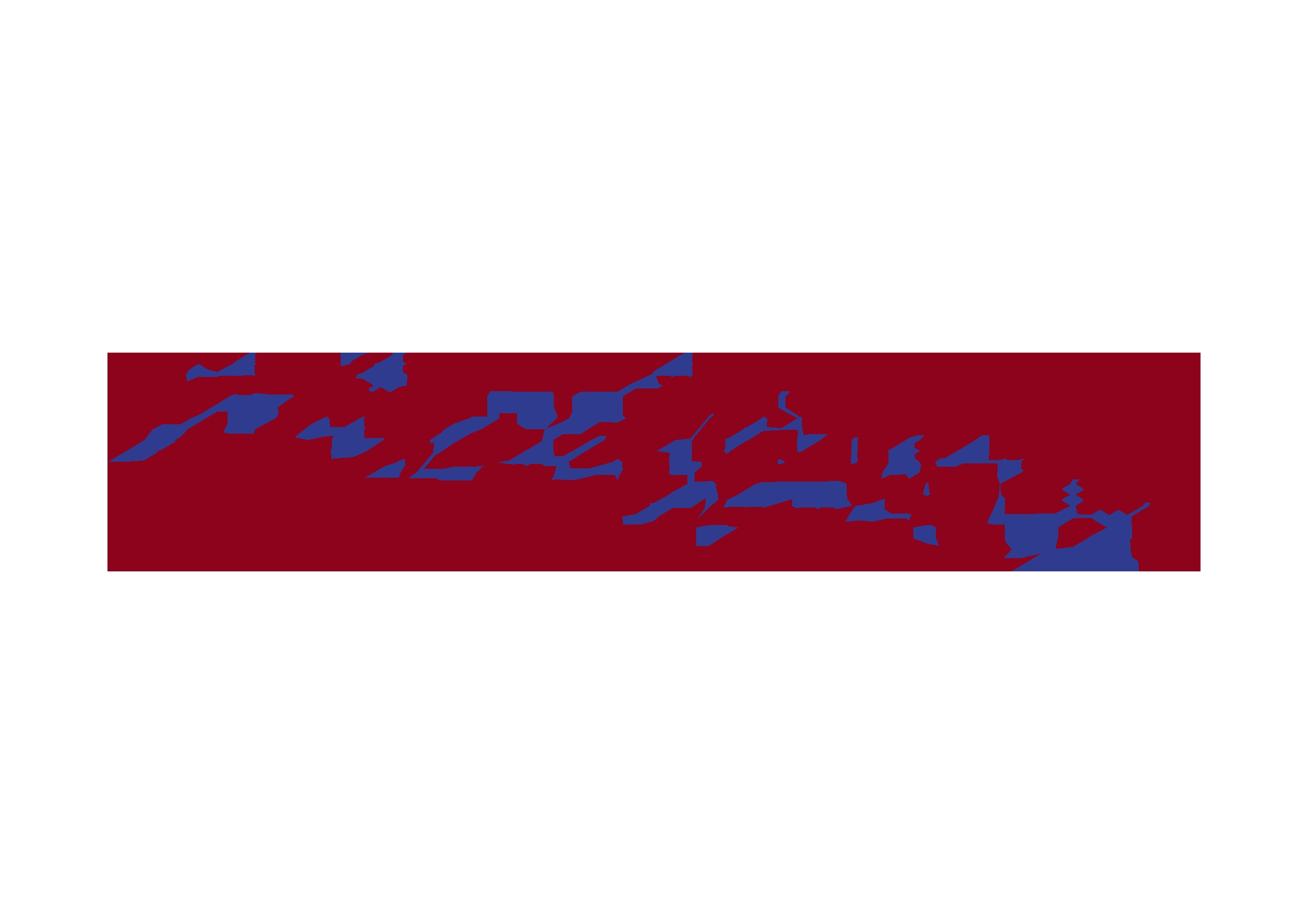}}
\end{subfigure}
\caption{Portion of the domain (in blue), on the plane $y=0.5$, in which the condition $\nu_{sgs}/\nu < -0.3$ is satisfied for $Re=3000$ at $t=4$. (a) Isotropic dynamic model. (b) Anisotropic dynamic model (component $22$ of $\nu_{sgs}/\nu$).}
\label{fig:nuratio_neg}
\end{figure}

A similar behaviour is obtained if we consider the minimum values of the ratio $\nu_{\rm sgs}/\nu$ (figure \ref{fig:min_nuratio}): as for the maximum values, figure \ref{fig:min_nuratio-3000} refers to $Re=3000$ while figure \ref{fig:min_nuratio-6000} to $Re=6000$. We have not reported the curve relative to the Smagorinsky model since it presents peak values nearly equal to zero, because of the fact that the Smagorinsky model is purely dissipative and does not allow backscatter. The dynamic model (blue dotted curve) presents, in absolute value, smaller negative peak values with respect to the anisotropic model (solid curves). Also in this case, the difference between the peak values provided by the isotropic and the anisotropic dynamic models is greater in the $Re=3000$ case.
Even if the Smagorinsky model appears to be more dissipative with respect to the other LES models if we look at the density isosurfaces and at the dissipated energy profiles, its subgrid viscosity peak values are lower. The Smagorinsky model is, in some sense, less selective with respect to the isotropic and anisotropic dynamic models and introduces a smaller, but more distributed dissipation.
This last statement can be checked representing, at a fixed instant of time, the field of the ratio $\nu_{sgs}/\nu$ for the different turbulence models.
In figure \ref{fig:nuratio}, which refers to the $Re=3000$ simulation at $t=4$, we represent in red the portion of the mesh in which the ratio $\nu_{sgs}/\nu$ takes values greater than $0.3$ for the Smagorinsky model (figure \ref{fig:nuratio_a}), the isotropic dynamic model (figure \ref{fig:nuratio_b}) and the anisotropic dynamic model (figure \ref{fig:nuratio_c}). Notice that for the anisotropic dynamic model the $22$ component of $\nu_{sgs}/\nu$ is represented, while the other components present similar patterns. 
If we compare the fields provided by the  Smagorinsky model and by the isotropic dynamic model, we can easily see that the portion of the mesh, for which the condition $\nu_{sgs}/\nu > 0.3$ is verified, is much smaller in the isotropic dynamic model case. For the different components of $\nu_{sgs}/\nu$, when employing the anisotropic dynamic model, the condition $\nu_{sgs}/\nu$ is verified for a greater portion of the mesh with respect to the isotropic dynamic model case. However, the overall dissipation is smaller than in the Smagorinsky model case (see figure \ref{fig:dissipated_energy_3d}): this is due to the presence of  backscatter, which is absent in the Smagorinsky model case. If we look at figure \ref{fig:nuratio_neg}, where the part of the mesh in which the condition $\nu_{sgs}/\nu < -0.3$ is verified is represented in blue, we can notice that for the anisotropic dynamic model (figure \ref{fig:nuratio_b_neg})  this condition is satisfied for  a large portion of the mesh, indicating that this model   well reproduces the backscatter phenomenon in the shear layer at the interface between the two densities. Notice that the presence of consistent backscatter in shear layers has already been highlighted for example in \cite{piomelli:1991}. The isotropic dynamic model (figure \ref{fig:nuratio_a_neg}) introduces much less backscatter with respect to the anisotropic counterpart.

\begin{figure}[]
\centering
\begin{subfigure}[]{
      \label{fig:rpe-3000}
      \includegraphics[width=0.9\textwidth]{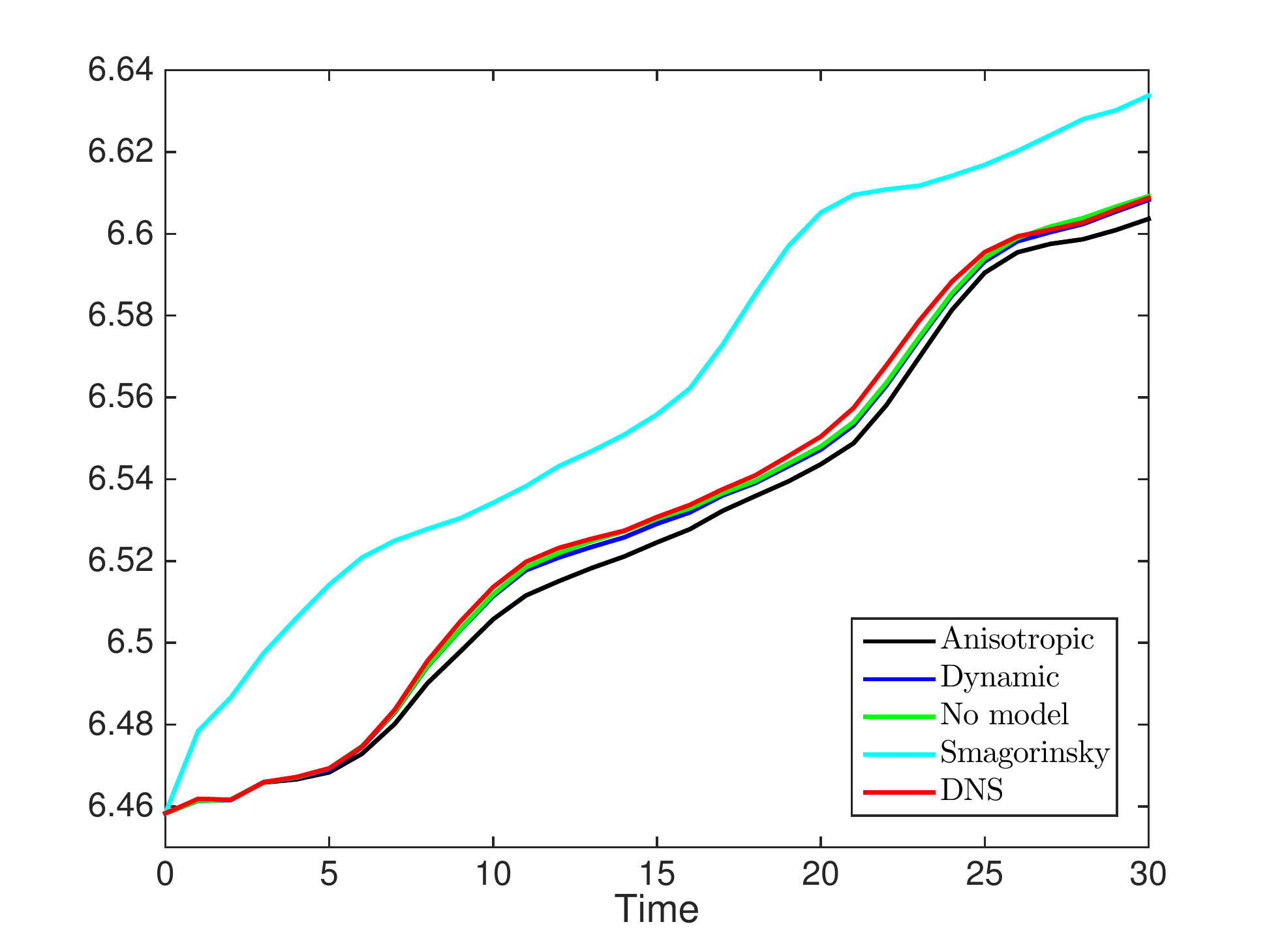}}
\end{subfigure} 
\begin{subfigure}[]{
        \label{fig:rpe-6000}
       \includegraphics[width=0.9\textwidth]{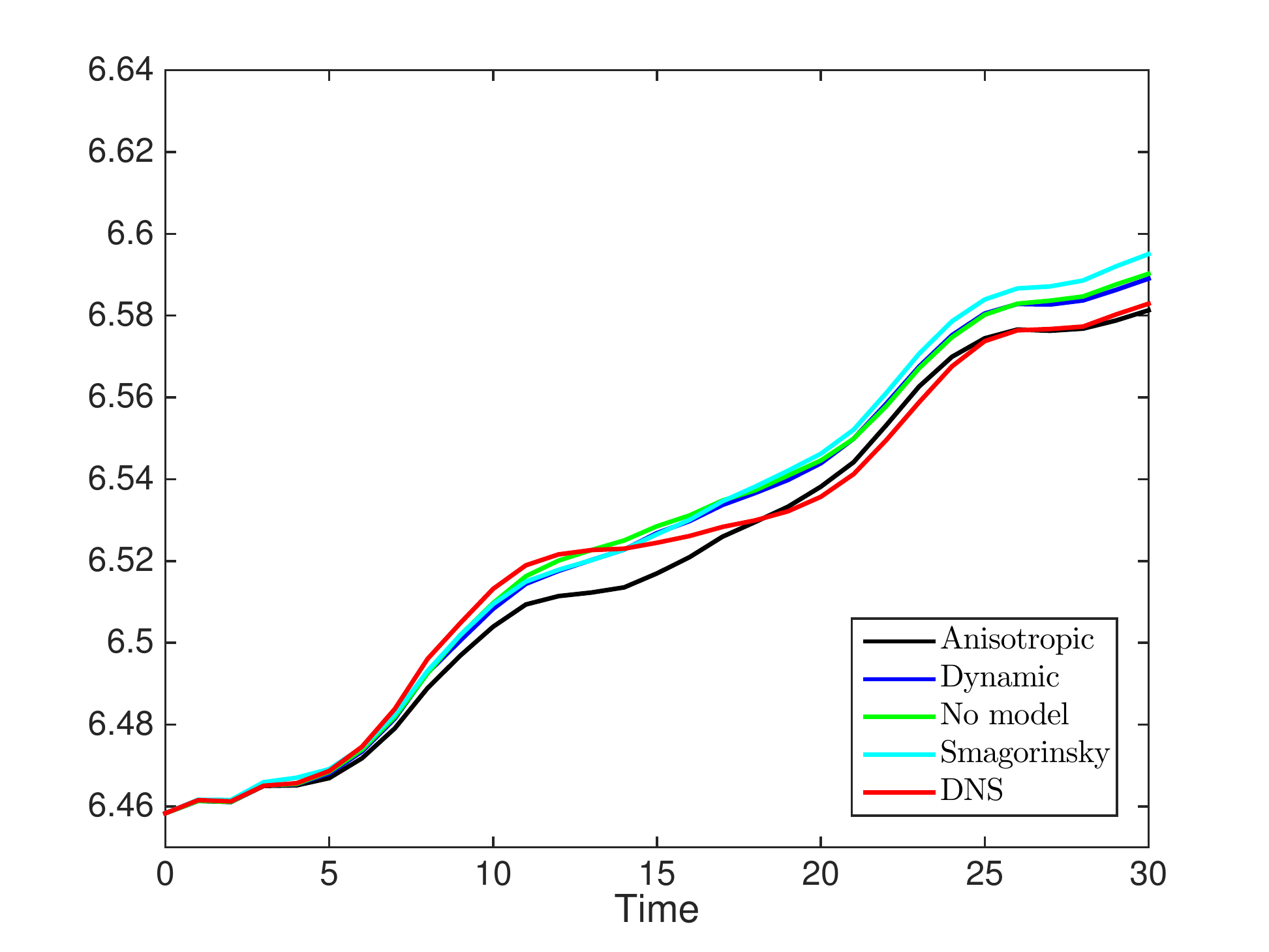}}
\end{subfigure}
\caption{Reference Potential Energy ($RPE$) as a function of time. (a) $Re=3000$. (b) $Re=6000$.}
\label{fig:rpe_3000_6000}
\end{figure}

As in \cite{ozgokmen:2007} and \cite{ozgokmen:2009}, we now perform a comparison between the different LES models employing the Reference Potential Energy ($RPE$), introduced at the beginning of the present section.
In figure \ref{fig:rpe_3000_6000}, we show the $RPE$ profiles as a function of time, obtained with the DNS and with different turbulence models. Figure \ref{fig:rpe-3000} refers to $Re=3000$, while figure \ref{fig:rpe-6000} to $Re=6000$. 
In the $Re=3000$ case, we can notice that the Smagorinsky model (cyan curve) has the worst performance, with a significant overestimation of the reference potential energy. The $RPE$ obtained with no model (green curve) and with the isotropic dynamic model (blue curve) are similar to each other and also quite similar to the DNS $RPE$. The anisotropic dynamic model (black curve) provides a slight underestimation of the $RPE$ profile. 

If we consider the $Re=6000$ case, we can notice that, after an initial slight underestimation of the DNS $RPE$, starting from $t > 15$, the $RPE$ provided by the anisotropic dynamic model (black curve) is very similar to the DNS $RPE$ (red curve), while the $RPEs$ provided by the no-model (green curve) and the isotropic dynamic model (blue curve) are very close and greater than the DNS $RPE$. The introduction of the Smagorinsky model (cyan curve) leads to a deterioration of the results in term of $RPE$ with respect to the no-model simulation, with an overestimation of the $RPE$. 

\begin{figure}
\centering
\includegraphics[width=0.5\textwidth]{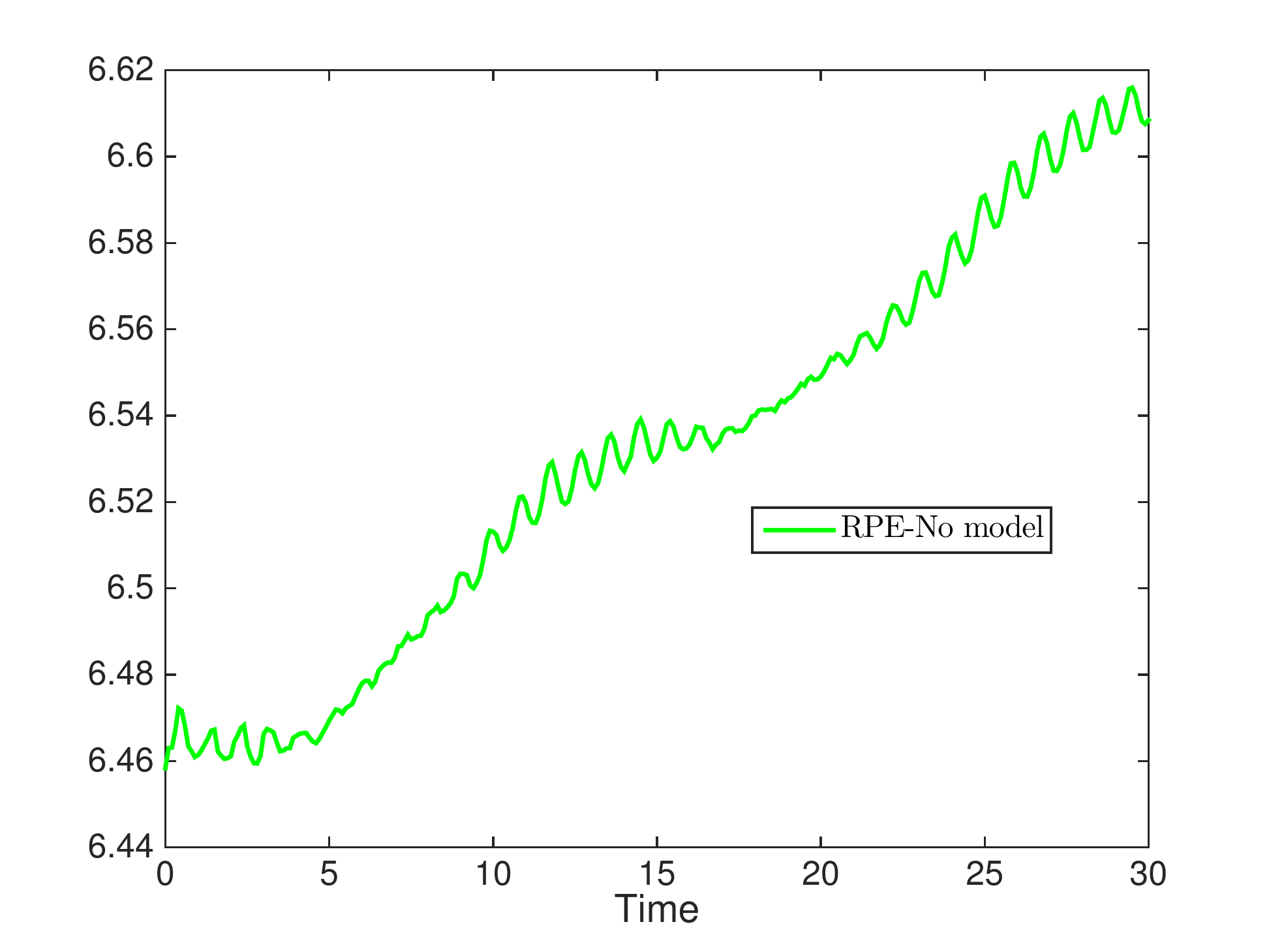}
\caption{$RPE$ as a function of time for $Re=3000$, obtained from the no model LES sampling the solution each $0.1$ non-dimensional time unit.}
\label{fig:nomod_rpe}
\end{figure}

Notice that the $RPE$ results presented in figure \ref{fig:rpe_3000_6000}, are obtained plotting the $RPE$ every non-dimensional time unit. If we present the plot of the $RPE$, for example obtained with the no model LES in the $Re=3000$ case, representing the $RPE$ every $0.1$ non-dimensional time units, we can notice that small-period oscillations appear (see figure \ref{fig:nomod_rpe}). 
We can easily see that the period of these oscillations is approximately equal to $1$ non-dimensional time unit. It can also be observed that the time employed by an acoustic wave to cover the length of the domain from $0$ to $L$ and then back from $L$ to $0$ is given by:
\begin{equation}
T_{2L} = 2 L Ma = 2 L 0.1 = 1,
\end{equation} 
in non-dimensional time units. As a consequence, we believe that the small period oscillations are caused by the fact that we are considering slightly compressible simulations 
in which acoustic perturbations can be induced by the abrupt start of the lock exchange test. Indeed,
these small period oscillations are instead absent in the incompressible simulations available in the literature.
The acoustic perturbations have probably an influence on the computation of the probability density function $P(\rho)$, employed in order to derive the $RPE$. We believe however that the presence of these small-period oscillations does not affect the general behaviour of the $RPE$, so that the comparison of the results obtained with different turbulence models in terms of $RPE$ is significant.

Concluding, the Smagorinsky model has the worst performance, with respect to both the instantaneous fields and the more quantitative diagnostics, as the dissipated energy and the $RPE$. 
Considering the other turbulence models, no model is able to correctly reproduce the large increase in the dissipated energy which is present in the DNS results. However, the anisotropic dynamic model is the one that, in particular in the $Re=6000$ case, appears to provide the best results in terms of the dissipated energy.
Concerning the $RPE$, in the $Re=3000$ case, the no-model and the isotropic dynamic model $RPEs$ are quite similar and also similar to the DNS one. The anisotropic dynamic model provides a slight underestimation of the $RPE$ in the $Re=3000$ case and also at the beginning of the $Re=6000$ simulation, but it appears to give the best results, especially in the $Re=6000$ case, toward the end of the simulation.

\section{Conclusions and future perspectives}
\label{conclu}
We have carried out three-dimensional simulations of non-Boussinesq gravity currents in the lock-exchange configuration. 
First,  three-dimensional DNS in the non-Boussinesq regime  were performed at two different Reynolds numbers. 
In the LES experiments we have employed the Smagorinsky model, the isotropic dynamic model and the anisotropic dynamic model, together with a no model run.
The considered diagnostics (instantaneous density and $Q$-criterion fields, time evolution of the dissipated energy and of the Reference Potential Energy) allow to confirm the fact that the Smagorinsky model is  too dissipative. Concerning the other models, the dynamic models (isotropic and anisotropic) seem to provide the best results in terms of dissipated energy and Reference Potential Energy, with slightly better results provided by the anisotropic dynamic model. 

Concerning the possible future developments, a first goal is the implementation of new turbulence models for variable density flows, based on the work \cite{germano:2014} and on the preliminary
\textit{a priori} analysis reported in \cite{bassi:2018a}, in the same configuration as   employed for the LES carried out here, in order to see if an improvement of the results can be obtained with respect to the results obtained with the turbulence models already tested. 

From a more numerical point of view, we also plan to implement a semi-implicit time integration scheme, following \cite{giraldo:2013}, \cite{tumolo:2015} and \cite{dumbser:2016}, in order to improve the computational efficiency, in particular when considering very low Mach number flows. In order to improve  efficiency, we   then plan to perform $p-$adaptive simulations following \cite{tumolo:2013}, \cite{tumolo:2015} and \cite{tugnoli:2017}.

\appendix
\section{Subgrid-scale models}
\label{section:sgsmod}
All the subgrid models we describe are built
on the Boussinesq hypothesis \cite{sagaut:2006}, which affirms that the energy
transfer mechanism from the resolved scales to the subgrid scales is analogous
to the molecular diffusion phenomenon, represented by the diffusion
term. A term which has the same mathematical structure as the molecular
diffusion term is then introduced in the NS equations to model the subgrid
stress tensor. More precisely, similarly to the molecular viscosity, which
is related to the magnitude of the molecular diffusion, a subgrid-scale viscosity
is introduced in order to set the magnitude of the energy transfer
between the resolved and the unresolved scales.
\subsection{The Smagorinsky model}
\label{section:smag_model}
In  the Smagorinsky subgrid model, first introduced in \cite{smagorinsky:1965}, the deviatoric part of the subgrid stress tensor $\tau_{ij}$ in (\ref{filteq}) is modelled by a scalar turbulent viscosity $\nu^{\rm sgs}$:
\begin{subequations}\label{eqn:nu_smag}
\begin{align}
 &  \tauij -\frac{1}{3}\tau_{kk} \deltaij = - \frho \nu^{{\rm sgs}} \fSij^d,
\label{eqn:nu_smag:tauij}
 \\
& \nu^{{\rm sgs}} =  C_S^2\Delta^2 |\fS|,
\label{eqn:nu_smag:nu}
\end{align}
\end{subequations}
where $C_S$ is the Smagorinsky constant (equal to $0.1$ in the present work), $|\fS|^2 = \dfrac{1}{2}\fSij\fSij$ and $\Delta$ is the filter scale. 
The isotropic part of the subgrid stress tensor can be modelled as:
\begin{equation}\label{eqn:taukk_smag}
 \taukk = C_I \frho\Delta^2 |\fS|^2.
\end{equation}
The subgrid temperature flux (\ref{eqn:Qj_sgs}) is set proportional to the resolved temperature gradient:
\begin{equation}\label{eqn:Qj_smag}
 Q_i^{{\rm sgs}} = - \frac{1}{\PR^{{\rm sgs}}} \frho \nusgs \de_i\fT,
\end{equation}
where $\PR^{{\rm sgs}}$ is a subgrid Prandtl number. 
The term $\tau(u_i,u_k,u_k)$ in the subgrid turbulent diffusion flux $J_j^{{\rm sgs}}$ (\ref{eqn:Jj_sgs}) is neglected by analogy with RANS leading to:
\begin{equation}\label{eqn:Jj_smag}
 J_i^{{\rm sgs}} \approx 2\fu_k\tau_{ik} + \fu_i\taukk.
\end{equation}

\subsection{The Germano dynamic model}
\label{section:germano_model}
In the  Germano dynamic model \cite{germano:1991}, the terms $C_S$ and $C_I$ of the Smagorinsky model are no more chosen \textit{a-priori} for the whole domain, but are  computed dynamically as  functions of the resolved field. The deviatoric part of the stress tensor is the same as in the Smagorinsky model:
\begin{equation}
  \tauij -\frac{1}{3}\tau_{kk} \deltaij = - \frho C_S \Delta^2 |\fS| \fSij^d.
\label{eqn:nu_iso:tauij}
\end{equation}
The coefficient $C_S$ is dynamically computed by introducing a test filter operator $\hat{\cdot}$ associated to a spatial scale $\hdelta$ which is larger than the spatial scale $\Delta$ related to the filter $\bbar{\cdot}$. 
A Favre filter (see equation (\ref{eq:favre_filter})), denoted with $\breve{\cdot}$, is associated to the test filter through the following Favre decomposition:
\begin{equation}\label{eqn:testfavre_decomp}
 \wh{\rho f} = \wh{\rho} \breve{f}.
\end{equation}
If the test filter $\hat{\cdot}$ is applied to the momentum equation (\ref{filteq-momentum}) we obtain:
\begin{equation}
\de_t \left( \wh{\rho} \breve{u}_i \right) + \de_j \left(\wh{\rho} \breve{u}_i \breve{u}_j\right) 
+ \de_i \widehat{p} - \de_j
\widehat{\sigma}_{ij} 
= - \de_j \left( \wh{\tau}_{ij} + \Lij \right),
\label{eq:momentum-test-averages}
\end{equation}
where
\begin{equation}\label{eqn:leo_qdm}
 \Lij = \wh{\frho\fu_i\fu_j} - \hfrho\hfu_i\hfu_j
\end{equation}
is the Leonard stress tensor. 
If we assume that the deviatoric part of the term at the right-hand side of equation (\ref{eq:momentum-test-averages}) can be modelled using an eddy viscosity model and we employ a least square approach as described in \cite{abba:2015}, we obtain for the Smagorinsky constant $C_S$ the following expression:
\begin{equation}\label{eq:dynamic-C_S}
 C_S = \dfrac{ \Lij^d \mR_{ij}}{\mR_{kl}\mR_{kl}},
\end{equation}
where $\mR_{kl} =  \wh{\frho \Delta^2 |\fS| \fS^d_{kl}} -
  \hfrho \hdelta^2 |\hfS| \breve{\fS^d}_{kl}$. 

The same dynamic procedure is applied to the isotropic component of the subgrid stress tensor: 
\begin{equation}\label{eqn:taukk_iso}
 \taukk = C_I \frho\Delta^2 |\fS|^2,
\end{equation}
where the $C_I$ coefficient is determined by:
\begin{equation}\label{eqn:dynamic-C_I}
 C_I= \dfrac{ \Lkk}{ \hfrho \hdelta^2 |\hfS|^2 - \wh{\frho \Delta^2 |\fS|^2}}.
\end{equation}

A similar approach is proposed for the subgrid terms in the energy equation. 
The subgrid heat flux is defined as:
\begin{equation}\label{eqn:Qj_iso}
Q_i^{\rm sgs} = - \frho \Delta^2 |\fS| C_Q \partial_i \widetilde{T}.
\end{equation}
The coefficient $C_Q$ in equation (\ref{eqn:Qj_iso}) is then obtained as:
\begin{equation} \label{eq:dynamic-C_Q}
C_Q = \frac{\Li^Q \mR^Q_i}{\mR^Q_k \mR^Q_k},
\end{equation}
with $\mR^Q_i = \wh{ \frho \Delta^2 |\fS| \de_i \fT} - \wh{\frho} \hdelta^2 |\hfS| \de_i \hfT$ and $ \mL_i^Q = \wh{\frho\fu_i\fT} - \hfrho\hfu_i\hfT$ temperature Leonard flux.

Concerning the subgrid turbulent diffusion flux, the term $\tau(u_i,u_k,u_k)$ in equation (\ref{eqn:Jj_sgs}) is not neglected as in the Smagorinsky model but a scale similarity model is assumed and this term is approximated as a subgrid kinetic energy flux:
\begin{equation}
\tau(u_i,u_k,u_k) \approx \frho \widetilde{u_i u_k u_k} -\frho \widetilde{u}_i\widetilde{u_k u_k}.
\label{eq:subgrid_kin_flux}
\end{equation}
The subgrid kinetic energy flux in equation (\ref{eq:subgrid_kin_flux}) $\tau(u_i,u_k,u_k)$ is then modeled as a function of the gradient of the resolved kinetic energy:
\begin{equation}
\tau(u_i,u_k,u_k) = - \frho \Delta^2 |\fS| C_J \de_i \left(\frac{1}{2}\widetilde{u}_k\widetilde{u}_k \right).
\label{eq:res_kin_en_grad}
\end{equation}
Introducing the kinetic energy Leonard flux:
\begin{equation}
\mL_i^J = \widehat{\frho \widetilde{u}_i\widetilde{u}_k \widetilde{u}_k} - \wh{\frho}\breve{\widetilde{u}}_i \breve{\widetilde{u}}_k \breve{\widetilde{u}}_k,
\end{equation}
the value of the constant $C_J$ is computed as:
\begin{equation}
C_J = \frac{\mL_i^J\mR_i^J}{\mR_k^J \mR_k^J},
\label{eq:C_J_dyn}
\end{equation}
where $R_i^J = \widehat{\frho \Delta^2 |\fS| \de_i \left(\frac{1}{2}\widetilde{u}_k \widetilde{u}_k\right)} - \widehat{\frho} \widehat{\Delta^2} |\hfS| \de_i \left( \frac{1}{2} \breve{\widetilde{u}}_k \breve{\widetilde{u}}_k \right).$
Considering equations (\ref{eqn:Jj_sgs}) and (\ref{eq:res_kin_en_grad}), the final expression for the subgrid turbulent diffusion flux is:
\begin{equation}
 J_i^{{\rm sgs}} = - \frho \Delta^2 |\fS| C_J \de_i\left(\dfrac{1}{2}\fu_k\fu_k\right) + 2\fu_k\tau_{ik} + \fu_i\taukk,
\label{eqn:Jj_iso}
\end{equation}
where $C_J$ is given by equation (\ref{eq:C_J_dyn}). 
It is important to point out that all the dynamic coefficients are averaged over each element in order to avoid numerical instabilities; moreover, since the dynamic model allows backscattering, a clipping procedure is applied to ensure that the total dissipation, resulting from both the viscous and the subgrid stresses, is positive.

\subsection{The anisotropic dynamic model}
\label{section:aniso_model}
We describe here the anisotropic dynamic model introduced in \cite{abba:2001} and extended to the compressible flows case in \cite{abba:2015}. The main characteristic of this anisotropic model is that it overcomes the limitation of the alignment between the subgrid flux tensors and the corresponding gradients, introducing tensorial proportionality coefficients between the two. 

If we consider in particular the momentum equation, the subgrid stress tensor $\tau_{ij}$ is assumed to be proportional to the strain rate tensor through a fourth order symmetric tensor as:
\begin{equation}
 \tauij  = - \frho \Delta^2 |\fS| \mB_{ijrs} \fS_{rs}.
 \label{eq:aniso_model}
\end{equation}
The coefficient $\mB_{ijrs}$ is dinamically computed following the procedure described in \cite{abba:2015}. $\mB_{ijrs}$ is rewritten as:
\begin{equation}
\mB_{ijrs} = \sum_{\alpha,\beta=1}^{3} \mC_{\alpha \beta} a_{i\alpha} a_{j \beta} a_{r\alpha} a_{s\beta},
\label{eq:Bijrs_def}
\end{equation}
where $a_{ij}$ is a rotation tensor and $\mC_{\alpha\beta}$ is a second order symmetric tensor.
As in \cite{abba:2015} we set $a_{ij}=\delta_{ij}$ and the following expression is obtained for $C_{ij}$:
\begin{equation}
\mC_{ij} = \frac{\Lij}{\left( \widehat{\frho \Delta^2 |\fS|\fS_{ij}} -\hfrho \hdelta^2 |\hfS|  \breve{\fS}_{ij} \right)}
\end{equation}
and 
\begin{equation}
\tauij  = - \frho \Delta^2 |\fS|\mC_{ij} \fS_{ij},
\label{eq:tau_ij_aniso}
\end{equation}
where no summation over the repeated indices is employed in the last formula.

As for the isotropic version of the dynamic model, the coefficients $\mC_{ij}$ are averaged over each element and a clipping procedure is applied in order to ensure that the total dissipation is positive. Notice that, in the anisotropic version of the dynamic model, the deviatoric and isotropic parts of $\tau_{ij}$ are modeled together. 

The dynamic procedure employed for the subgrid-scale stress in the momentum equation is applied also for the subgrid terms in the energy equation. 

The subgrid heat flux is expressed as:
\begin{equation}
Q_i^{\rm sgs} = - \frho \Delta^2 |\fS|\mB_{ir}^Q \partial_r \widetilde{T}, 
\label{eq:aniso_model_heatflux}
\end{equation}
with $\mB_{ir}^Q$ symmetric tensor. If $\mB_{ir}^Q$ is diagonal, considering the reference frame defined by the tensor $a$, the following equation is obtained:
\begin{equation}
\mB_{ir}^Q = \sum_{\alpha=1}^{3} \mC_\alpha^Q a_{i\alpha} a_{r\alpha}.
\label{eq:tensor_aniso_heatflux}
\end{equation}
Notice that the coefficients $\mC_\alpha^Q$ can be computed via the dynamic procedure as in \cite{abba:2015}, obtaining: 
\begin{equation}
\mC_\alpha^Q = \frac{a_{i\alpha}\mL_i^Q}{a_{r\alpha} \left( \widehat{\frho \Delta^2 |\fS|\de_r\widetilde{T}} -\hfrho \hdelta^2 |\hfS|  \de_r\breve{\widetilde{T}} \right)}.
\end{equation}
The subgrid kinetic energy flux, given by equation (\ref{eq:subgrid_kin_flux}), is modeled as:
\begin{equation}
\tau(u_i,u_k,u_k) = -\frho \Delta^2 |\fS| \mB_{ir}^J \de_r \left(\frac{1}{2}\widetilde{u}_k \widetilde{u}_k \right),
\end{equation}
with:
\begin{equation}
\mB_{ir}^J = \sum_{\alpha=1}^3 \mC_\alpha^J a_{i\alpha} a_{r\alpha}.
\end{equation}
The coefficient $\mC_\alpha^J$ is dinamically computed as:
\begin{equation}
\mC_\alpha^J = \frac{ a_{i\alpha} \mL_i^J}{\mM_\alpha},
\end{equation}
where 
\begin{equation}
\mM_\alpha = a_{r\alpha} \left( \widehat{\frho \Delta^2 |\fS| \de_r \left(\frac{1}{2}\widetilde{u}_k \widetilde{u}_k\right)} - \hfrho \hdelta^2 |\hfS| \partial_r \left( \frac{1}{2} \breve{\widetilde{u}_k}\breve{\widetilde{u}}_k \right) \right).
\end{equation}
Finally the subgrid turbulent diffusion flux takes the following form:
\begin{equation}
 J_i^{{\rm sgs}} = - \frho \Delta^2 |\fS| \mB_{ir}^J \de_r\left(\dfrac{1}{2}\fu_k\fu_k\right) + 2\fu_k\tau_{ik} + \fu_i\taukk.
 \label{eq:sgs_turb_diff_flux_aniso}
\end{equation}

\section{Numerical method}
\label{section:dgmeth}
The filtered Navier-Stokes equations are spatially discretized by the Discontinuous Galerkin finite elements method. The DG approach is analogous to that described in \cite{giraldo:2008}. In particular the Local Discontinuous Galerkin (LDG) method is chosen for the approximation of the second order viscous terms (see \cite{arnold:2002}, \cite{bassi:1997},  \cite{castillo:2000}, \cite{cockburn:1998}). 
The Navier-Stokes equations (\ref{filteq}) are rewritten in compact form and introducing an auxiliary variable $\bmG$, so that
\begin{eqnarray}
\label{eq:csv_auxvar}
\de_t \bU + \Div \bF^{{\rm c}}(\bU) &=&  \Div \bF^{{\rm v}}(\bU,\bmG) \nonumber\\
 &-&\Div \bF^{{\rm sgs}}(\bU,\bmG) + \bS   \\
 \bmG &-& \nabla{\boldsymbol \varphi} = 0,\nonumber
\end{eqnarray}
where $\bU=[\frho\,,\frho\wt{\bu}^T,\frho\fe ]^T$ are the prognostic
variables, ${\boldsymbol \varphi} = [\wt{\bu}^T,\fT]^T$ are the variables
whose gradients enter the viscous fluxes~(\ref{eq:constitutive-Favre}),
as well as the turbulent ones  and $\bS$ represents the source terms.
The fluxes in (\ref{eq:csv_auxvar}) are written in the following compact form:
\[
\bF^{{\rm c}} = \left[ \frho\wt{\bu},\frho\wt{\bu}\otimes\wt{\bu}+\bbar{p}\mathcal{I},\frho\wt{h}\wt{\bu} \right]^T,
\]
\[
\bF^{{\rm v}} = \left[ 0, \wt{\sigma}, \wt{\bu}^T \wt{\sigma}-\wt{\bq} 
\right]^T,
\]
and
\[
\bF^{{\rm sgs}} = \left[ 0, \tau,  \frac{1}{(\gamma-1)\MA^2} \mathbf{Q}^{{\rm sgs}}
 +\frac{1}{2}\left( 
 \mathbf{J}^{{\rm sgs}} - \tau_{kk} \wt{\bu}
 \right) 
 \right],
 \]

\[
\bS = \left[
 0,
 \frho \mathbf{f},
\frho \mathbf{f} \cdot \wt{\bu}
\right].\]
Here, $\tau$, $\mathbf{Q}^{{\rm sgs}}$ and $\mathbf{J}^{{\rm sgs}}$
are given by~(\ref{eqn:nu_smag}), (\ref{eqn:Qj_smag})
and~(\ref{eqn:Jj_smag}), respectively, for the Smagorinsky model, by~(\ref{eqn:nu_iso:tauij}), (\ref{eqn:Qj_iso})
and~(\ref{eqn:Jj_iso}) for the isotropic dynamic model and by (\ref{eq:tau_ij_aniso}), (\ref{eq:aniso_model_heatflux}) and (\ref{eq:sgs_turb_diff_flux_aniso}) for the anisotropic dynamic model.  

To define the space discretization, a tessellation $\mT_h$ of $\Omega$ into tetrahedral elements $K$ is introduced such that $\Omega =
\bigcup_{K\in\mT_h} K$ and $K\cap K'=\emptyset$ and the finite element space is defined as:
\begin{equation}\label{eqn:mV_def}
\mV_h = \left\{ v_h \in L^2(\Omega): v_h|_K \in \mathbb{P}^p(K), \,
\forall K\in\mT_h \right\},
\end{equation}
where $p$ is a nonnegative integer and $\mathbb{P}^p(K)$ denotes the
space of polynomial functions of total degree at most $p$ on $K$. 
For
each element, the outward unit normal on $\partial K$ will be denoted
by $\bn_{\partial K}$. Given $d$ the spatial dimension of the problem, the numerical solution is now defined as
$(\bU_h,\bmG_h)\in(\,(\mV_h)^{(2+d)}\,,\,(\mV_h)^{4\times d}\,)$ such that,
$\forall K\in\mT_h$, $\forall v_h\in\mV_h$, $\forall
\br_h\in(\mV_h)^d$,

\begin{subequations}
\label{eq:DG-space-discretized}
\begin{align}
\displaystyle
\frac{d}{dt}\int_K \bU_h v_h\,d\bx
& \displaystyle
- \int_K \bF^{\rm c}(\bU_h)\cdot\nabla v_h\, d\bx
\\[3mm]
& \displaystyle
+ \int_{K} (\bF^{\rm v} - \bF^{\rm sgs})(\bU_h,\bmG_h) \cdot \nabla v_h\, d\bx
\\[3mm]
& \displaystyle
+ \int_{\partial K} \bF^{\rm{c},*}(\bU_h,\bU_h^+)\cdot \bn_{\partial K} v_h\,
d\sigma
\\[3mm]
& \displaystyle
- \int_{\partial K} (\bF^{\rm v}-\bF^{\rm sgs})({\boldsymbol \varphi}^*,\bmG^*)\cdot \bn_{\partial K} v_h\,
d\sigma
= \int_K \bS v_h \,d\bx,\nonumber 
\\[3mm] \displaystyle
\int_K \bmG_h \cdot \br_h \,d\bx
& \displaystyle
+ \int_K {\boldsymbol \varphi_h}\nabla\cdot\br_h\, d\bx
\\[3mm]
& \displaystyle
- \int_{\partial K} {\boldsymbol \varphi}^* \bn_{\partial
K}\cdot\br_h \, d\sigma = 0, \nonumber 
\end{align}
\end{subequations}
where $\bU_h=\left[ \rho_h\,,\rho_h\bu_h\,,\rho_he_h \right]^T$ is the DG approximation of the solution into the element $K$,
${\boldsymbol \varphi}_h=\left[ \bu_h\,,T_h \right]^T$ are the quantities for which the gradient is computed, $\bF^{\rm{c},*}$ is the numerical flux associated to the convective term, while ${\boldsymbol \varphi}^*$ and $\bmG^*$ are the numerical fluxes for ${\boldsymbol \varphi}_h$ and $\bmG_h$. The numerical fluxes are responsible for the coupling between elements. Notice that the symbols $\bU_h$ and $\bU_h^+$, when appearing as arguments of the numerical flux functions, assume the meaning described in the following. Consider the element $K$, the edge $e \in K$ and a point $\xi \in e$, we have:
\begin{subequations}
\begin{align}
&\bU_h = \bU_h(\xi^{int(K)}) = \lim_{\mathclap{{x \rightarrow \xi},{x \in K}}} \bU_h(x), \\
&\bU_h^+ = \bU_h(\xi^{ext(K)})= \lim_{\mathclap{{x \rightarrow \xi},{x \notin K}}} \bU_h(x).
\end{align}
\label{eq:xi_int_xi_ext}
\end{subequations}

If only piecewise constant basis functions are considered, the weak formulation given by equations (\ref{eq:DG-space-discretized}) defines implicitly the standard low order Finite Volume approaches. From this viewpoint, the Discontinuous Galerkin method can be seen as an extension to arbitrary order of accuracy of the Finite Volume method and, as a consequence, the well known exact and approximate Riemann solvers developed in the Finite Volume context can be successfully employed for the construction of the numerical flux $\bF^{\rm{c},*}$ associated to the convective terms (see \cite{toro:2009} for a comprehensive review of Riemann solvers). 
We have employed the exact Godunov Riemann solver implemented as in \cite{gottlieb:1988}.

As already anticipated, the definition of the numerical flux for the viscous and subgrid terms is realized by means of the Local Discontinuous Galerkin (LDG) method. In particular we have extensively employed the method proposed in \cite{bassi:1997}, obtaining:

\begin{subequations}
\begin{align*}
&\bmG^* = \frac{1}{2} \left(\bmG_h +\bmG_h^+\right), \\
&\boldsymbol \varphi* = \frac{1}{2} \left(\boldsymbol{\varphi}_h  +\boldsymbol {\varphi}_h^+ \right),
\end{align*}
\end{subequations}
where the symbols have the same meaning as in equations (\ref{eq:xi_int_xi_ext}).

Concerning the choice of the polynomial basis, on each element, the unknowns are expressed in terms of an orthogonal basis, yielding what is
commonly called a modal DG formulation. All the integrals in equations (\ref{eq:DG-space-discretized}) are evaluated using quadrature formulae
from~\cite{cools:2003}, which are exact for polynomial orders up to $2p$. This results in a diagonal mass matrix in the time derivative term of~(\ref{eq:DG-space-discretized}) and simplifies the computation of $L^2$ projections to be introduced in connection with the LES filters. 

Notice that the fact that the quadrature formula employed for the numerical integration are exact for polynomial degrees up to $2p$ guarantees the exact integration of the mass matrix. However, the same quadrature rules are insufficient for the exact integration of the convective flux terms, which present more complex non-linearities. This may lead, expecially for high polynomial degrees ($p>4$), to severe aliasing errors which tipically cause instabilities and crashing of the computation due to the generation of strong unphysical oscillations, which are the cause of the violation of the constitutive laws \cite{beck:2016}.
Notice that, in \cite{beck:2016}, it is pointed out that, also in the case of low order polynomial degrees ($p\leq2$) and mid range polynomial degrees, such as $p=3$, for which stability is not a problem due to the presence of sufficient numerical dissipation, which counteracts the aliasing effect, the aliasing terms may negatively influence the solution: the interaction between discretization, aliasing and subgrid model dissipation gives indeed rise to inaccurate results. 

A possible solution to the aliasing issue is the so called polynomial de-aliasing, first introduced in \cite{kirby:2003}: the numerical quadrature precision is increased such that the convective flux integrals can be evaluated exactly with respect to the machine precision. Notice that, as discussed in \cite{beck:2016}, the introduction of a de-aliasing procedure is important because in this way the physically based turbulence models can better carry out their role of modeling the missing subgrid-scale physics and are not relegated to countering the numerical instabilities arising from aliasing errors. 

In the following the filter operators $\bbar{\cdot}$ and $\wh{\cdot}$, introduced in section \ref{turbulence_model}, will be explicitly defined in the context of the DG finite elements method. In particular, following a VMS approach,  filter operators are identified with an $L^2$ projection, as suggested e.g. in \cite{collis:2002b},  \cite{collis:2002a}, \cite{vanderbos:2007} and \cite{abba:2015}. 
Given a subspace $\mV\subset L^2(\Omega)$, let
$\Pi_{\mV}:L^2(\Omega)\to\mV$ be the associated projector defined by
\[
\int_\Omega \Pi_{\mV}u\,v\, d\bx =
\int_\Omega u\,v\, d\bx, \qquad \forall u,v \in\mV,
\]
where the integrals are evaluated with the same quadrature rule used
in~(\ref{eq:DG-space-discretized}). For $v\in L^2(\Omega)$, the filter
$\bbar{\cdot}$ is now defined by
\begin{equation}
\bbar{v} = \Pi_{\mV_h}v.
\label{eq:filter-bar}
\end{equation}
Notice that the application of this filter is built in the
discretization process and equivalent to it. Therefore,
once the discretization of equations (\ref{eq:csv_auxvar}) has
been performed, only $\bbar{\cdot}$ filtered
quantities are computed by the model. 
To define the test filter, we introduce
\begin{equation}\label{eqn:mVhat_def}
\wh{\mV}_h = \left\{ v_h \in L^2(\Omega): v_h|_K \in
\mathbb{P}^{\wh{p}}(K), \, \forall K\in\mT_h \right\},
\end{equation}
where $0\leq\wh{p}<p$, and we let, for $v\in L^2(\Omega)$,
\begin{equation}
\wh{v} = \Pi_{\wh{\mV}_h}v.
\label{eq:filter-hat}
\end{equation}
By our previous identification
of the $\bbar{\cdot}$  filter and the discretization,
the quantities $\frho$, $\frho\wt{\bu}$ and $\frho\fe$
can be identified with
  $\rho_h$, $\rho_h \bu_h $ and $\rho_he_h,$ respectively.
  Therefore they belong to $\mV_h,$ for which an orthogonal
  basis is employed by the numerical method.
  As a result, the computation of $\wh{\rho_h}$, $\wh{\rho_h\bu_h}$
and $\wh{\rho_he_h}$ is straightforward and reduces to zeroing the
last coefficients in the local expansion.
Assuming that the analytic solution is defined in some infinite
dimensional subspace of $L^2$, heuristically, $\mV_h\subset L^2$
is associated to the scales which are represented by the model, while
$\wh{\mV}_h\subset\mV_h\subset L^2$  is associated to
the spatial scales  well resolved by the numerical approximation. 


The spatial scales $\Delta$ and $\hdelta$ associated with the two filters~(\ref{eq:filter-bar}) and~(\ref{eq:filter-hat}) can be computed as:
\begin{equation}
\Delta = \left(\frac{\Delta_x\Delta_y\Delta_z}{N_p}\right)^{1/3}, \quad \hdelta = \left(\frac{\Delta_x\Delta_y\Delta_z}{N_{\widehat{p}}}\right)^{1/3},
\end{equation}
where $N_p$ and $N_{\widehat{p}}$ are the number of degrees of freedom per element associated to the polynomial degrees $p$ and $\widehat{p}$, respectively and $\Delta_x$, $\Delta_y$ and $\Delta_z$ are equivalent grid spacings computed as in equations (\ref{eq:resolutions}).  The filter scales are, as a consequence, piecewise polynomial functions in space. 

 \section*{Acknowledgements} 
 This paper is part of the first author's PhD thesis work.
 We are happy to acknowledge the continuous help of M. Restelli and M.Tugnoli with the application of the FEMILARO code. Several comments by F. Denaro and M.V. Salvetti have also been very useful
 to improve the presentation of some results.
  The results of this research have been achieved using the 
computational resources made available   at CINECA (Italy) by the LISA high performance computing project
 {\it DECLES: Large Eddy Simulation of Density Currents and Variable Density Flows, HPL13PJ6YS}.
 
\bibliographystyle{plain}
\bibliography{lock_exchange}

\end{document}